\pdfoutput=1

\documentclass[ amsmath, amssymb, aps, prd, twocolumn, lengthcheck, sort&compress, showpacs, superscriptaddress, nofootinbib ]{revtex4-1}

\usepackage{graphicx}
\usepackage{dcolumn}
\usepackage{amssymb}
\usepackage{mathrsfs}
\usepackage{amsmath}
\usepackage{comment}
\usepackage{amsbsy}
\usepackage{dsfont}
\usepackage{units}
\usepackage{nicefrac}

\DeclareFontFamily{OT1}{pzc}{}
\DeclareFontShape{OT1}{pzc}{m}{it}{<-> s * [1.10] pzcmi7t}{}
\DeclareMathAlphabet{\mathpzc}{OT1}{pzc}{m}{it}

\newcommand{\Slash}[1]{\ooalign{\hfil/\hfil\crcr$#1$}}

\usepackage[usenames]{color}
\usepackage{colortbl}

\usepackage[colorlinks=true,linkcolor=blue,citecolor=blue,urlcolor=blue]{hyperref}

\definecolor{lred}{rgb}{1,0.90,0.7}

\begin{document}
\title{Delta and Omega  masses  in a three-quark covariant Faddeev approach}
\author{Helios \surname{Sanchis-Alepuz}}
\email{helios.sanchis-alepuz@uni-graz.at}
\affiliation{Institute of Physics, University of Graz, Universit\"atsplatz 5, 8010, Graz, Austria}
\author{Gernot  \surname{Eichmann}}
\affiliation{Institut f\"{u}r Theoretische Physik I, Justus-Liebig-Universit\"at Giessen, D-35392 Giessen, Germany}
\author{Selym \surname{Villalba-Ch\'avez}}
\thanks{present address: Max-Planck-Institut f\"ur Kernphysik,\\ Saupfercheckweg 1, D-69117 Heidelberg, Germany}
\affiliation{Institute of Physics, University of Graz, Universit\"atsplatz 5, 8010, Graz, Austria}
\author{Reinhard \surname{Alkofer}}
\affiliation{Institute of Physics, University of Graz, Universit\"atsplatz 5, 8010, Graz, Austria}

\date{\today}

\begin{abstract}
We present the solution of  the Poincar\'e-covariant Faddeev equation for the
$\Delta(1232)$ and $\Omega(1672)$ baryons. The covariant
structure of the corresponding baryon amplitudes and their decomposition in
terms of internal spin and orbital angular momentum is explicitly derived.
The interaction kernel is truncated to a rainbow-ladder dressed-gluon
exchange such that chiral symmetry and its dynamical breaking are correctly
implemented. The resulting physical masses agree reasonably with experiment
and their evolution with the pion mass compares favorably with lattice
calculations. Evidence for the non-sphericity of the $\Delta-$resonance is
discussed as well.
\end{abstract}



\maketitle

\section{Introduction}

Quantum Chromodynamics (QCD) was built upon the understanding of hadrons via
the quark model and the discovery of color. However, despite the success of QCD
predictions at high energies, which led to the consensus that QCD provides a
correct description of strong interactions, the calculation of hadron
properties developed at a slower pace. This is due to the fundamental
non-perturbative character of bound-state phenomena, and it is further complicated
by the low-energy enhancement of strong interactions.

Hadrons, as bound states of quarks and gluons, are described in continuum QCD
by (generalized) Bethe-Salpeter equations (BSEs)
\cite{Salpeter:1951sz,Faddeev:1960su,Taylor:1966zza,Boehm:1976ya,Loring:2001kv} 
which rely upon the Green functions
of the theory. This, in principle, necessitates a solution of the  infinite tower of
coupled  Dyson-Schwinger equations  (DSEs)~\cite{Alkofer:2000wg,Fischer:2006ub,Chang:2011vu}. Any feasible
numerical procedure thus requires a symmetry-preserving truncation,
but once such a truncation is performed, the approach provides a fully
consistent quantum-field theoretical framework for the study of hadron
properties.

The aforementioned approach has 
a longstanding and successful history in the investigation of meson properties; see,
e.g., \cite{Maris:2005tt,Maris:2006ea,Krassnigg:2009zh} and references therein. For baryons, however, the
relativistic bound-state description is considerably more involved and
computationally demanding. The reason is that the presence of a third
quark enlarges the momentum phase space, and the relativistic spin
structure \cite{Carimalo:1992ia,Eichmann:2009en} of the three-quark bound-state
amplitude is much more complicated as well. A successful simplification
of the problem starts from a covariant Faddeev equation~\cite{Taylor:1966zza,Boehm:1976ya}
but immediately reduces its complexity by
treating the two-body scattering matrix in a separable expansion: by
considering quark-quark correlations, called diquarks, as dominant,
the three-quark equation is simplified to a quark-diquark BSE
\cite{Hellstern:1997pg,Oettel:1998bk,Bloch:1999ke}. For a collection of recent
results using the quark-diquark approach see, e.g., Refs.\
\cite{Eichmann:2007nn,Eichmann:2008ae,Eichmann:2008ef,Nicmorus:2008vb,Nicmorus:2010sd,Eichmann:2010je,Nicmorus:2010mc}.

A step forward was taken in Refs.~\cite{Eichmann:2009zx,Eichmann:2009qa,Alkofer:2009jk,Eichmann:2009en} where,
for the first time, the full
three-body Poincar{\'e}-covariant  Faddeev equation  for the nucleon was solved. Here, the
interaction kernel is truncated to a rainbow-ladder dressed-gluon exchange between two
of the quarks, and irreducible three-quark contributions are neglected. The
nucleon mass reported in these works is comparable to
lattice data and, surprisingly, also very close to the quark-diquark result.
The calculation has been recently improved in Ref.~\cite{Eichmann:2011vu}.

In the family of baryon resonances, the $\Delta(1232)$ isoquadruplet plays a
special role, owing to its  high production cross section which makes it one of
the best studied resonances. Moreover, it is the lightest baryon resonance and
possesses the same valence-quark content as the nucleon, and thus constitutes
an excellent system to understand excitations in QCD. It is also the lightest
known particle with spin $\nicefrac{3}{2}$.
At present, due to numerical limitations, such covariant three-body bound-state calculations are
performed only for baryons with equal-mass valence quarks. This is the case for
the $\Delta$ quadruplet assuming isospin symmetry. 
Another spin-$\nicefrac{3}{2}$ ground-state baryon is the
$\Omega(1672)$: it is a pure strange state and therefore allows to study the quark-mass dependence of
$\Delta$ properties.

In the present work we extend the techniques described in~\cite{Eichmann:2009qa,Eichmann:2009en,Eichmann:2011vu}
to the case of the $\Delta$ and $\Omega$ resonances in view of computing their masses and bound-state amplitudes.
We analyze the importance of the different partial-wave contributions
and thereby demonstrate that the $\Delta$ and $\Omega$ baryons are not pure s-wave
states. This represents a first step towards the understanding of baryon
deformation from sphericity. A more complete description of baryon  structure
will require to study the  contributions of different internal spin and angular
momentum structures to physical observables such as, {\it e.g.}, electromagnetic
form factors or the electromagnetic $\gamma N\Delta$-transition.

The paper is organized as follows. In Section \ref{sec:Faddev_eq} we review the
Poincar\'e-covariant Faddeev approach to baryons and the rainbow-ladder truncation
of the interaction kernel, and we describe the main features of the covariant
decomposition of the Faddeev amplitude for spin-$\nicefrac{3}{2}$ baryons. In
Section \ref{sec:Results} we present our results for the $\Delta$ and $\Omega$
masses and compare with lattice data and experimental values. We also discuss
the internal angular momentum composition of the amplitudes. 
Technical details about the partial-wave decomposition and the
numerical implementation are described in Appendices \ref{sec:basis} and
\ref{sec:Numerics}, respectively. Our calculations
are performed in Euclidean momentum space and Landau gauge QCD;
details about the conventions used in this paper can be found, e.g., in Ref.~\cite{Eichmann:2011vu}.

\section{Faddeev Equation and Rainbow-ladder truncation}\label{sec:Faddev_eq}

Baryons are described in QCD by the three-quark (six-point) Green function
which characterizes the state both on- and off-shell. On the baryon mass shell
the Green function has a pole, and a Laurent expansion around this pole allows
to derive a relativistic three-body bound-state equation (see
Ref.\ \cite{Loring:2001kv} for a pedagogical discussion).
That equation reads
\begin{equation}
\Psi =\pmb{K}_3\,G_0^{(3)}\,\Psi\,, \qquad \pmb{K}_3 =
 \pmb{K}_3^{irr} + \sum_{a=1}^3 \pmb{K}^{(a)}_{(2)}\,,
\end{equation}
where $\Psi$ represents the baryon's bound-state amplitude and
$G_0^{(3)}$ is the product of three dressed quark propagators.
The three-body kernel $\pmb{K}_3$ includes all possible correlations among the
three quarks: it comprises a three-body irreducible contribution
$\pmb{K}_3^{irr}$ as well as
the sum of the three two-body irreducible interactions with a
spectator quark $a$, $\pmb{K}^{(a)}_{(2)}$.
The bound-state amplitude $\Psi$ satisfies
the physical normalization condition
\begin{equation}\label{eq:normalization}
 \mathcal{N}\bar{\Psi}\left[\frac{d}{dP^2}
 \left(\pmb{K}_3^{-1}-G_0^{(3)}\right)
 \right]\Psi=1\,,
\end{equation}
where $\bar{\Psi}$ is the charge-conjugated amplitude and $P$ the total baryon momentum.

The structure of the bound-state amplitude depends on the baryon of interest.
It is the product of color, flavor and spin parts. For spin-$\nicefrac{1}{2}$
baryons, the spin part is a rank-four Dirac tensor, with three Dirac indices
representing the valence quarks and the fourth index the
bound state. A spin-$\nicefrac{3}{2}$ particle is described by a
Rarita-Schwinger field so that in this case the spin part of the bound-state amplitude
is a mixed tensor with four Dirac indices and one Lorentz index.
Moreover, it depends on the three quark momenta $p_1$, $p_2$ and $p_3$ which
can be conveniently reexpressed in terms of the total momentum $P$ and two
relative momenta $p$ and $q$:
     \begin{equation}
     \begin{array}{rl@{\quad}rl}
        p &= (1-\zeta)\,p_3 - \zeta p_d\,, &  p_1 &=  -q -\dfrac{p}{2} + \dfrac{1-\zeta}{2} P\,, \\[0.25cm]
        q &= \dfrac{p_2-p_1}{2}\,,         &  p_2 &=   q -\dfrac{p}{2} + \dfrac{1-\zeta}{2} P\,, \\[0.25cm]
        P &= p_1+p_2+p_3\,,                &  p_3 &=   p + \zeta  P\,,
     \end{array}
     \end{equation}
where $p_d=p_1+p_2$. The total momentum is constrained by $P^2=-M^2$, with $M$ being the baryon mass.
$\zeta$ is a free momentum partitioning parameter which we will choose to be
$\zeta=\nicefrac{1}{3}$. This choice maximizes the accessible bound-state
mass range with respect to the analytic structure of the quark-propagator~\cite{Eichmann:2009zx}
and also allows for a tremendous simplification of the
bound-state equation's solution method (see Appendix \ref{sec:Numerics}).

     \begin{figure*}
     \begin{center}
     \includegraphics[width=.95\textwidth]{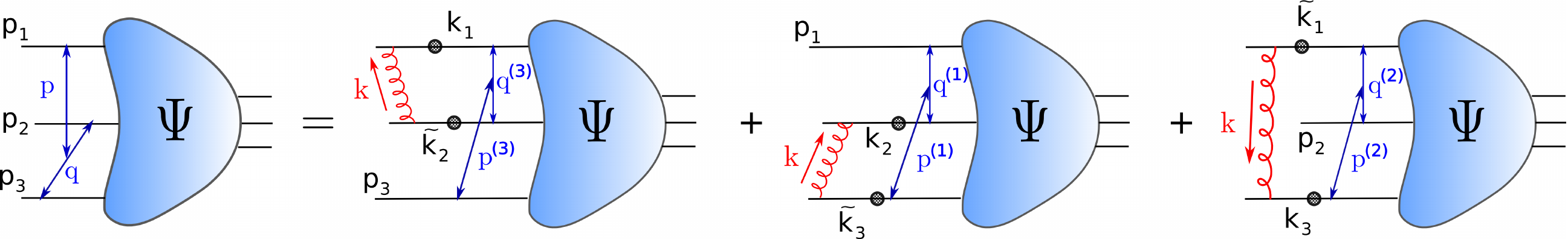}
     \caption{Diagrammatical representation of  the Faddeev equation in
     rainbow-ladder truncation.}\label{fig:faddeev-eq}
     \end{center}
     \end{figure*}

The success of the quark-diquark approach to baryon properties 
supports the idea that quark-quark correlations dominate the binding of
baryons. Consequently, we neglect the three-body irreducible contribution
$\pmb{K}_3^{irr}$. Then, considering the above discussion about the
bound-state amplitude's tensorial structure and kinematics, one arrives at the
covariant Faddeev equation  for a spin-$\nicefrac{3}{2}$ baryon (see Fig.~\ref{fig:faddeev-eq}):
\begin{widetext}
\begin{eqnarray}\label{eq:faddeev_eq}
\Psi_{\alpha\beta\gamma\delta}^\mu(p,q,P) &=&
\int_k  \left[ K_{\beta\beta'\gamma\gamma'}(k)S_{\beta'\beta''}(k_2)
S_{\gamma'\gamma''}(\tilde{k}_3)
\Psi_{\alpha\beta''\gamma''\delta}^\mu(p^{(1)},q^{(1)},P)\right.\nonumber\\
&&\quad \left. +K_{\gamma\gamma'\alpha\alpha'}(k)S_{\gamma'\gamma''}(k_3)
S_{\alpha'\alpha''} (\tilde{k}_1)
\Psi_{\alpha''\beta\gamma''\delta}^\mu(p^{(2)},q^{(2)},P)
\right. \nonumber\\
&&\quad  \left. + K_{\alpha\alpha'\beta\beta'}(k)S_{\alpha'\alpha''}(k_1)
S_{\beta'\beta''}(\tilde{k}_2)
\Psi_{\alpha''\beta''\gamma\delta}^\mu(p^{(3)},q^{(3)},P)\right] \, ,
\end{eqnarray}
\end{widetext}
where we have already restricted ourselves to the case where the two-quark
kernel depends only on the exchange (i.e., gluon) momentum $k$. The quark propagators $S$
depend on the internal quark momenta $k_i=p_i-k$ and $\tilde{k}_i=p_i+k$.
The internal relative momenta are
\renewcommand{\arraystretch}{1.2}
\begin{equation}\label{internal-relative-momenta}
\begin{array}{l@{\quad}l@{\quad}l}
p^{(1)} = p+k,& p^{(2)} = p-k,& p^{(3)} = p,\\
q^{(1)} = q-k/2,& q^{(2)} = q-k/2, & q^{(3)} = q+k.
\end{array}
\end{equation}

The solution of the Faddeev equation is greatly simplified if one expresses
the Faddeev amplitude in terms of a basis $\tau_{\alpha\beta\gamma\delta}^i\,^\mu(p,q,P)$,
\begin{equation}
\label{eq:basis_decomposition}
\Psi_{\alpha\beta\gamma\delta}^\mu(p,q,P)=\sum_i f_i(p^2,q^2,\{z\})\,
\tau_{\alpha\beta\gamma\delta}^i\,^\mu(p,q,P) \,,
\end{equation}
where the expansion coefficients $f_i$ depend on the
five Lorentz-invariant variables
\begin{equation}\label{momentum-invariants}
p^2,\quad q^2,\quad z_0=\widehat{p_T}\cdot\widehat{q_T},\quad z_1=\hat{p}\cdot\hat{P},\quad
z_2=\hat{q}\cdot\hat{P}\,.
\end{equation}
Here, a hat denotes a normalized four-vector and the subscript $T$
a transverse projection with respect to the total momentum $P$.


In principle one is free to choose any possible basis~$\tau^i$.
For the sake of physical interpretation it is convenient to perform a
partial-wave decomposition and classify the basis elements with respect to
their quark-spin and relative orbital angular momentum content in the baryon's rest frame. The explicit
derivation of such a basis is quite involved~\cite{SanchisAlepuz:2010in} and
presented in Appendix~\ref{sec:basis}. The most significant aspects of this construction are:
\begin{itemize}
 \item It is independent of any approximation in the Faddeev equation.
 \item Only Poincar\'e covariance as well as parity invariance are needed to
 construct the basis.
 \item For positive-parity, positive-energy (particle) spin-$\nicefrac{3}{2}$
 baryons the basis consists of $128$ elements.
 \item The basis includes all possible spin and orbital angular momentum
 values, namely, $\mathpzc{s}=\nicefrac{1}{2}\,,\nicefrac{3}{2}$ and
 $\ell=0\,,1\,,2\,,3$.
\end{itemize}

Once the Poincar\'e-covariant structure of the spin-$\nicefrac{3}{2}$ baryon amplitude is determined,
the elements in the Faddeev equation that remain to be specified are the
two-quark interaction kernel and the quark propagator.
They are related by the axial-vector Ward-Takahashi identity (AXWTI)
which, in pseudo-scalar meson studies, guarantees the correct implementation of chiral symmetry and its
spontaneous breaking. In particular,
it implies that pions are massless in the chiral limit~\cite{Delbourgo:1979pt,Finger:1980dw,Fomin:1984tv,Munczek:1991jb,Munczek:1994zz,Maris:1997hd,Holl:2004fr}.
The simplest $qq$ interaction kernel that satisfies the AXWTI is a dressed-gluon ladder exchange:
\begin{equation}\label{eq:RL_kernel}
\mathcal{K}_{\alpha\alpha'\beta\beta'}(k) = Z_2^2 \,\frac{4\pi\alpha(k^2)}{k^2} \,
 T^{\mu\nu}_k \,\gamma^\mu_{\alpha\alpha'} \,\gamma^\nu_{\beta\beta'}\,,
\end{equation}
where $Z_2$ is the quark renormalization constant,
$T^{\mu\nu}_k = \delta^{\mu\nu}-\hat{k}^\mu \hat{k}^\nu$ is a transverse projector with respect to
the gluon momentum $k$,
and $\alpha(k^2)$ is an effective interaction that defines the model input.

The quark propagator satisfies the quark Dyson-Schwinger equation whose interaction
kernel includes the dressed gluon propagator as
well as one bare and one dressed quark-gluon vertex.
Consistency with the AXWTI requires to use the same interaction~\eqref{eq:RL_kernel} in the
quark DSE:
\begin{equation}\label{eq:RL_QuarkDSE}
S^{-1}_{\alpha\beta}(p) = Z_2 \left( i\Slash{p} + m_0 \right)_{\alpha\beta}
+ \int_q \mathcal{K}_{\alpha\alpha'\beta'\beta}(k) \,S_{\alpha'\beta'}(q)\,,
\end{equation}
where $m_0$ is the bare current-quark mass that enters the equation as an input, and $k=q-p$.
Tracing the color structure yields a prefactor $\nicefrac{4}{3}$ in front of the DSE integral in~\eqref{eq:RL_QuarkDSE}
and $\nicefrac{2}{3}$ for the integral in the Faddeev equation~\eqref{eq:faddeev_eq}.
Eqs.~(\ref{eq:RL_kernel}--\ref{eq:RL_QuarkDSE}) define the rainbow-ladder truncation where the dressed quark-gluon vertex
is truncated to its vector part $\gamma^\mu$, and the combined non-perturbative dressing
of the gluon propagator and quark-gluon vertex is absorbed in the
effective interaction $\alpha(k^2)$ which has to be modeled.

For the effective coupling we adopt the ansatz introduced in Ref.~\cite{Maris:1999nt}
which has been successfully applied in many hadron studies.
It reproduces the logarithmic behaviour of
QCD's one-loop running coupling in the ultraviolet, and it is strong enough in
the infrared to enable dynamical chiral symmetry breaking and thereby generates a
dynamical constituent-quark mass:
\begin{equation}\label{eq:couplingMT}
\alpha(k^2) = \pi\eta^7\left(\frac{k^2}{\Lambda^2}\right)^2 \!\!
e^{-\eta^2\left(k^2/\Lambda^2\right)} +
\alpha_{UV}\left(k^2\right).
\end{equation}
The infrared part is characterized by one dimensionless parameter $\eta$
and an energy scale $\Lambda$. The 'ultraviolet term' reads
            \begin{equation}
                \alpha_\text{UV}(k^2) = \frac{2\pi\gamma_m \big(1-e^{-k^2/\Lambda_t^2}\big)}{\ln\,[e^2-1+(1+k^2/\Lambda^2_\mathrm{QCD})^2]}\,,
            \end{equation}
where $\gamma_m=12/(11N_C-2N_f)$ is the anomalous dimension of the  quark
propagator. We use $\gamma_m=12/25$ and $\Lambda_{QCD}=0.234$ GeV.  The scale
$\Lambda_t=1$ GeV is only introduced for technical reasons, its precise value has no
impact on the results. The infrared energy scale is fixed to the value
$\Lambda=0.72$ GeV in order to reproduce the pion decay constant
\cite{Maris:1999nt}.
Many meson observables, and here especially the ones related to
ground-state pseudoscalar and vector mesons, turn out to be
almost insensitive to the infrared width parameter $\eta$ around a central
value $\eta=1.8$~\cite{Krassnigg:2009zh},
and a similar observation holds for nucleon properties~\cite{Nicmorus:2010mc,Eichmann:2011vu}.

           \begin{table}
            \begin{tabular}{llll}
           \hline\hline
                                       & Nucleon~~~~~~~               & $\Delta$~~~~~~~  & $\Omega$  \\ \hline
           Physical~~~~~               & 0.94                         & 1.23             & 1.67  \\
           Q-DQ \cite{Nicmorus:2010mc} & 0.94                         & 1.28             & 1.77  \\
           Faddeev                     & 0.94 \cite{Eichmann:2011vu}  & 1.26             & 1.72  \\ \hline
            \end{tabular}
           \caption{Nucleon, $\Delta$ and $\Omega$ masses (in GeV) obtained from the
           Faddeev equation and, for comparison, in the quark-diquark (Q-DQ) approach~\cite{Nicmorus:2008vb,Nicmorus:2010mc}.
            In both cases, the parameter $\eta$ in Eq.~(\ref{eq:couplingMT}) is set to $\eta=1.8$. }\label{table_masses}
           \end{table}

\section{Results}\label{sec:Results}

           Having defined the input of the Faddeev equation,
           Eqs.~(\ref{eq:faddeev_eq}) and (\ref{eq:RL_QuarkDSE}) can now be
           solved consistently.
           The details of the calculation are given in Appendix~\ref{sec:Numerics}.
           From the solution of the Faddeev equation we obtain the
           bound-state mass and amplitudes, and the Faddeev amplitudes thereby provide
           information about the internal structure of the baryon.

           For the hadron mass calculation, the current-quark masses must be fixed  as
           well. Upon setting the value for the infrared scale $\Lambda$, the $u/d$ mass is adjusted to
           reproduce the physical pion mass. Since there is no purely strange pseudoscalar
           meson,  the $s$-quark mass is fixed to reproduce the kaon mass which would
           correspond  to a fictitious pseudoscalar $s\bar{s}$ state $m_{s\bar{s}}=690~$MeV.

           To analyze the sensitivity of our results to the parameters of the effective
           interaction, we have repeated our calculations for different values of
           $\eta$ in the range $1.6-2.0$. For the central value $\eta=1.8$, the
           resulting $\Delta$ mass at the physical pion mass, and the $\Omega$ mass
           at $m_{\pi}=690~$MeV, are shown in Table~\ref{table_masses}. They are in good
           agreement with the corresponding experimental values. For comparison, we have
           also included the respective mass results in the quark-diquark approach~\cite{Nicmorus:2008vb,Nicmorus:2010mc}.
           While the same value for the nucleon mass, $M_N=0.94$~GeV, is obtained both in the quark-diquark approach
           and through the Faddeev equation, the quark-diquark results for the spin-$\nicefrac{3}{2}$ baryon masses
           are larger. This solidifies the observation that
           a truncation of the Faddeev kernel to (ground-state) scalar and axial-vector diquark channels
           is sufficient to reproduce nucleon properties,
           whereas more structure is needed to describe spin-$\nicefrac{3}{2}$ baryons.
           Nevertheless, the results clearly demonstrate that diquarks capture an important part of the
           internal structure of baryons.


               \begin{figure}[t]
               \begin{center}
               \includegraphics[scale=0.57]{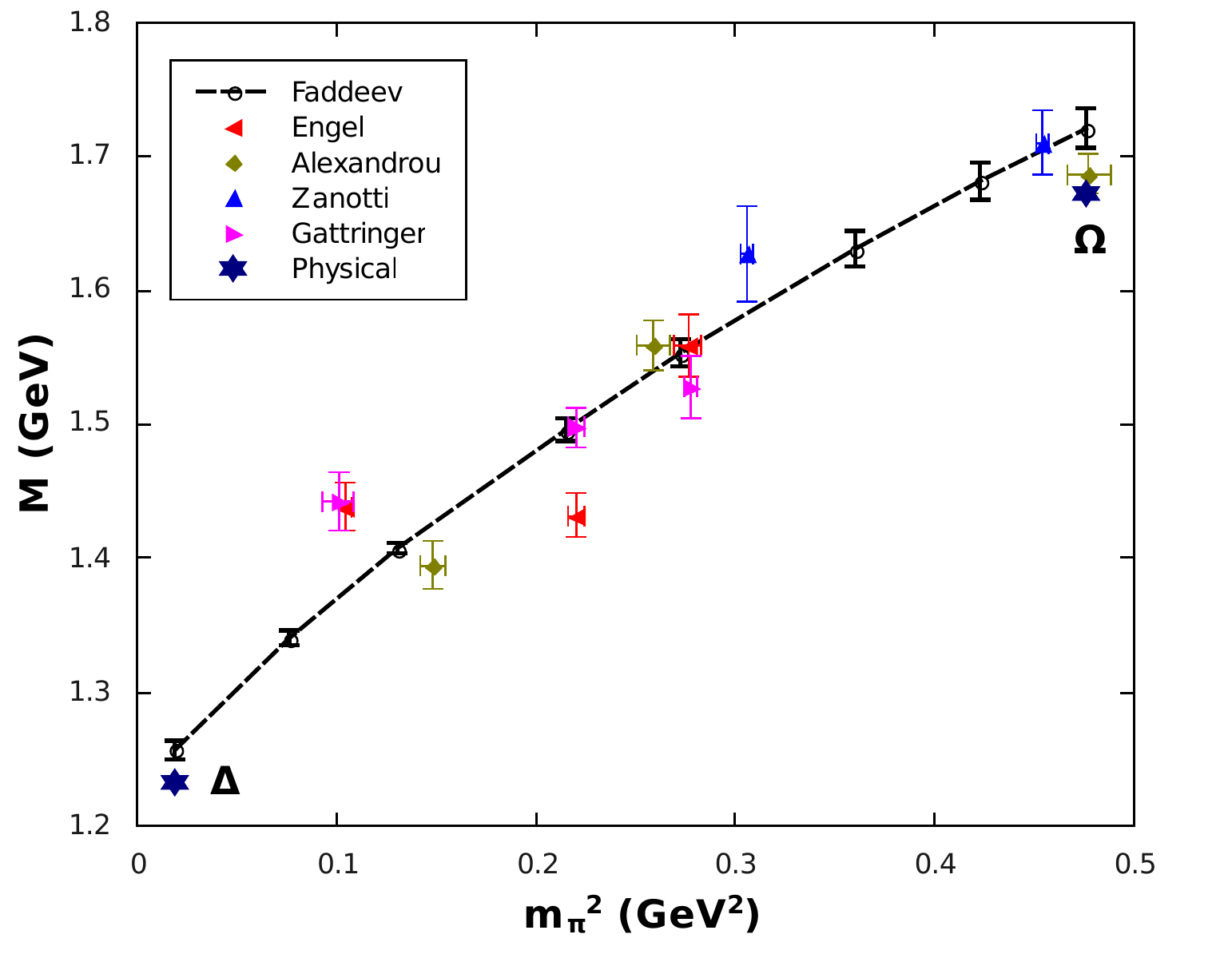}
               \caption{Evolution of $M_\Delta$ with $m_\pi^2$ compared to lattice data
               \cite{Alexandrou:2009hs,Engel:2010my,Zanotti:2003fx,Gattringer:2008vj}.
               Stars denote the experimental values of $\Delta$ and $\Omega$.
               Error bars in the Faddeev calculation correspond to a variation in the
               interaction width in the range $\eta\in 1.6 - 2.0$}\label{fig:mass_evolution}
               \end{center}
               \end{figure}

           It is interesting to study the relevance of interaction terms beyond the
           rainbow-ladder truncation. Meson studies show that pionic corrections
           (pseudoscalar-meson exchange among quarks, so-called resonant corrections) are
           attractive, while non-resonant contributions have a repulsive effect
           \cite{Fischer:2008wy,Fischer:2009jm,Chang:2009zb,Chang:2010hb}. Both effects seem to cancel at the
           non-perturbative level, so that the rainbow-ladder kernel gives an accurate
           description of the $q\bar{q}$ interaction in mesons. Moreover, pionic corrections
           are expected to be suppressed at high quark masses.

           The evolution of the $\Delta$-mass with the squared pion mass is shown in
           Fig.~\ref{fig:mass_evolution} and compared to lattice results. Here, the pion
           mass is obtained by solving the corresponding Bethe-Salpeter equation using the
           same interaction kernel upon varying the current-quark mass. We find a good
           agreement of the Faddeev-calculated masses with lattice data, especially at
           higher pion masses. If the analysis of meson results for effects beyond rainbow-ladder can be
           extended to the $qq$-interaction in baryons, this would indicate that a near
           cancelation of resonant and non-resonant contributions prevails up to 
           the strange-quark mass and beyond.

           
           Ultimately, such issues can only be clarified by overcoming the rainbow-ladder truncation
           as well as the restriction to quark-quark correlations in the Faddeev equation.
           We note, however, that the current-mass evolution of the $\Delta$~mass from the Faddeev equation
           is in qualitative agreement with that of the vector-meson mass obtained from its Bethe-Salpeter equation, 
           cf. Fig.~2 in Ref.~\cite{Nicmorus:2010mc}.
           Using the current-mass independent interaction defined in Eq.~\eqref{eq:couplingMT},
           both resulting curves overestimate the experimental values of $M_\Omega$ and $m_\Phi$ 
           at the strange-quark mass approximately by the same amount ($\sim 3\%$).
           In view of this, it could be
           misleading to attribute such discrepancies solely to corrections beyond rainbow-ladder
           without understanding the sensitivity of our calculations to the effective
           interaction $\alpha(k^2)$.  
           It would be desirable to compare results using
           different, lattice-inspired, model interactions; see also Ref.~\cite{Qin:2011dd}. 
           Such a study will be published elsewhere.

           There is another missing feature in the current-mass evolution of the
           $\Delta$ mass. The $\Delta$ resonance can decay into a nucleon via the emission
           of a pion. Such a decay would manifest itself in a non-analytical behaviour of
           the $\Delta$ mass as a function of the pion mass when the decay channel opens,
           \textit{i.e.} for $m_\pi\sim M_{\Delta}-M_N\sim 300~$MeV
           \cite{Allton:2005vm,Mader:2011zf}.
           Our truncation scheme does not provide a mechanism
           for the $\Delta$ to decay and thus we should obtain a higher $\Delta$ mass,
           shifted approximately $100~$MeV upwards, corresponding to the $\Delta$ decay
           width. Again, an analogous observation holds for the $\rho-$meson obtained
           through the Bethe-Salpeter equation~\cite{Maris:1999nt,Nicmorus:2010mc}.
           Therefore, we conclude that the model provides too much binding in the
           light pion-mass region where the decay channel would be opened in a full
           calculation.

           A first hint on the internal structure of the $\Delta$-baryon comes from the
           analysis of the relative importance of the different quark-spin and relative
           angular-momentum contributions. In Fig.~\ref{fig:amplitudes} we plot the
           dominant amplitudes in each partial-wave sector (see Appendix \ref{sec:basis}).
           It is clear that the $\mathpzc{s}=\nicefrac{3}{2}$, $\ell=0$ (\textit{i.e.}, s-wave)
           sector is dominant. However, p- and d-wave sectors are much larger in number of
           basis elements, thus even small values of individual basis elements can add up
           to a non-negligible contribution of a given sector and thereby cause deviations
           from sphericity. We obtain very similar
           results for the $\Omega$-baryon. In any case, a thorough understanding of the
           $\Delta$ and $\Omega$ internal structure requires the study of the different
           partial-wave sectors in physical observables, \textit{e.g.}, form factors.

               \begin{figure}[t]
               \begin{center}
               \includegraphics[scale=0.58]{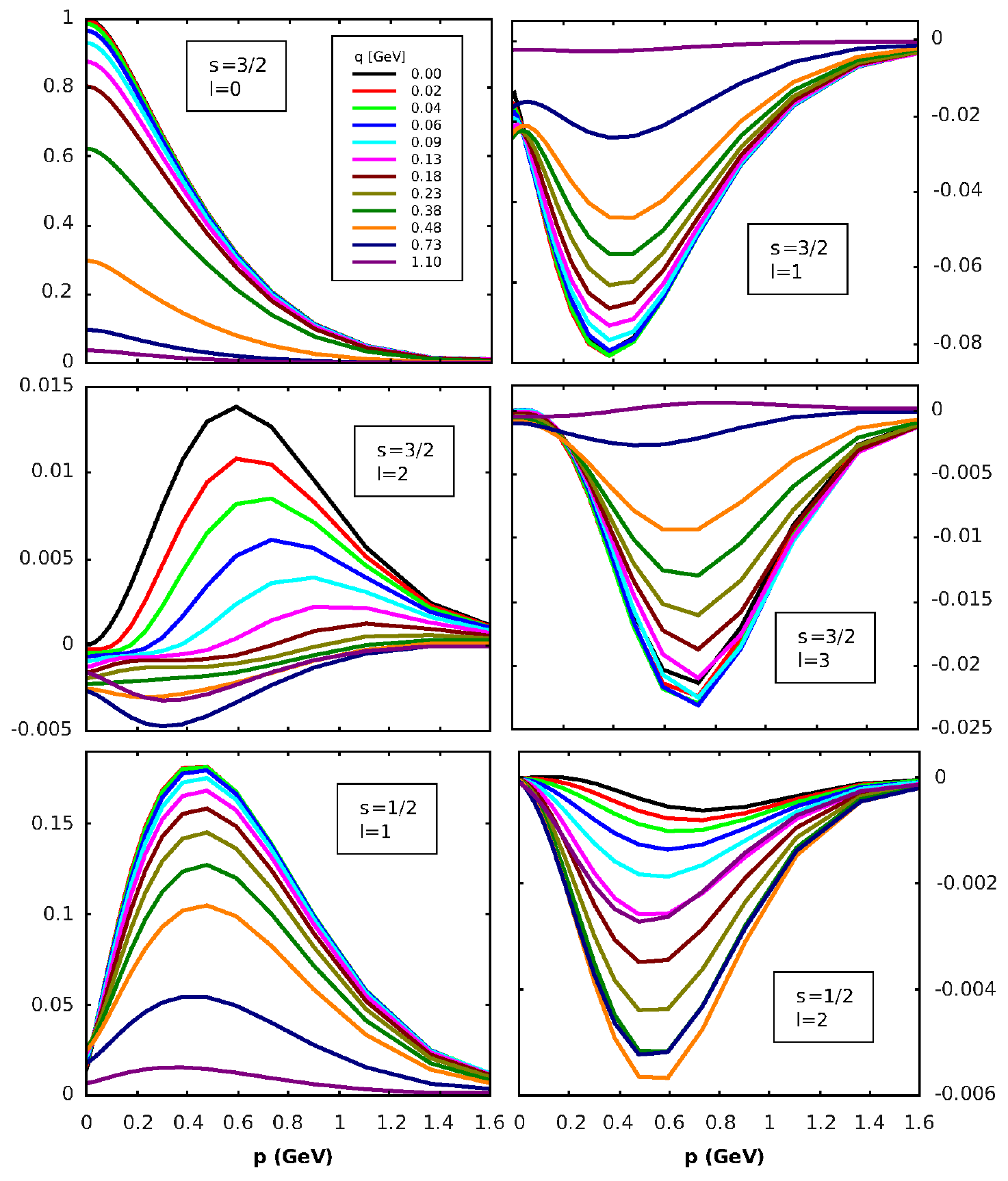}
               \caption{Zeroth Chebyshev moments of the coefficients $f_i$ in the basis
               expansion of Tables \ref{table_basis3_2} and \ref{table_basis1_2}.
               We plot the coefficients dominant in each $(\mathpzc{s},\ell)$-sectors as a function
               of $p$ and $q$. The normalization is chosen such that $f_1(p^2=0,q^2=0)=1$.
               The respective basis elements $\tau_{ij}^{\sigma,k}$ are from left to right, top to bottom:
               $\tau_{11}^{+1}$, $\tau_{13}^{+1}$, $\tau_{22}^{-2}$, $\tau_{33}^{-1}$, $\tau_{63}^{+1}$, $\tau_{74}^{+2}$.}
               \label{fig:amplitudes}
               \end{center}
               \end{figure}

\section{Conclusions}\label{sec:Summary}

We presented a fully Poincar\'e-covariant solution of the Faddeev equation for
the $\Delta$- and $\Omega$-baryons. The required quark propagators are obtained
by consistently solving the corresponding Dyson-Schwinger equations. For this we
employed a rainbow-ladder truncation of the interaction kernel, dressed by an
effective interaction which effectively  depends on a single parameter and is
fixed to reproduce the pion decay constant.

We obtain masses that are in good agreement with experimental results and their
evolution with the pion mass compares favorably with lattice results. We
concluded, however, that the effective interaction we used provides too much
binding in the low pion-mass region.

The contribution of the different quark-spin and relative orbital angular
momentum sectors was studied by na\"ively comparing the magnitude of the
corresponding dominant amplitudes. This shows an s-wave dominance, however with a
non-negligible p- and d-wave contribution. A better understanding of the
internal structure would come from the study of $\Delta$ and $\Omega$ form
factors in the present framework. Work in this direction is in progress.
\newline
~
\newline
~
\vspace{0.005 in}
\begin{flushleft}
\textbf{Acknowledgments}
\end{flushleft}
\vspace{0.005 in}

We are grateful to M.\ Blank, C.S.\ Fischer, A. Krassnigg, D.\ Nicmorus  and R.\ Williams for
helpful discussions. This work was supported by  the Austrian Science Fund FWF
under Projects No.\ P20592-N16,  Erwin-Schr\"odinger-Stipendium J3039,
and the Doctoral Program W1203; by the Helmholtz International Center for FAIR
within the LOEWE program of the State of Hesse; as well as  in part by the European Union
(HadronPhysics2 project ``Study of strongly-interacting matter'').

\begin{appendix}

\section{Covariant decomposition of the $\Delta-$baryon amplitude}\label{sec:basis}

            The spin part of the Faddeev amplitude for spin-$\nicefrac{3}{2}$ particles is
            a mixed tensor with four Dirac indices and one Lorentz index.
            The positive-parity and positive-energy subspace can be expanded in a basis
            with 128 elements $\tau^i_{\alpha\beta\gamma\delta}\,^\mu(p,q,P)$, cf.~Eq.~\eqref{eq:basis_decomposition},
            which can be
            classified with respect to their quark-spin and relative orbital angular
            momentum in the baryon's rest frame.
            The resulting basis is collected in Tables \ref{table_basis3_2}
            and \ref{table_basis1_2}, and in the following we will sketch its derivation.

    \renewcommand{\arraystretch}{1.5}

            \begin{table}[t]
            \begin{tabular}{ | c | c || c |} \hline\rule{0mm}{0.5cm}

            $\mathpzc{s}$ &  $\ell$  &   $\sqrt{5}\,\tau^{\sigma,1}_{1j}$ \\ [0.15cm]\hline\hline

            $\cellcolor{lred}\quad\nicefrac{3}{2}\quad$   &  \cellcolor{lred}$\quad 0\quad$    & \cellcolor{lred} $\sqrt{5}\,\textnormal{S}^g_{11}$ \\

            $\nicefrac{3}{2}$   &  $1$     &   $3\,\textnormal{S}^g_{12}+2\,(\textnormal{V}^r_{14}-\textnormal{V}^s_{13})$ \\

            $\cellcolor{lred}\nicefrac{3}{2}$   &  \cellcolor{lred}$1$     & \cellcolor{lred} $3\,\textnormal{S}^g_{13}+2\,\textnormal{V}^r_{11}$\\

            $\nicefrac{3}{2}$   &  $1$     & $3\,\textnormal{S}^g_{14}+2\,\textnormal{V}^s_{11}$ \\[0.1cm]
            \hline\hline\rule{0mm}{0.5cm}

            $\mathpzc{s}$           &  $\ell$  &   $\frac{1}{\sqrt{3}}\,\tau^{\sigma,1}_{2j}$   \\ [0.15cm]   \hline\hline
            $\quad\nicefrac{3}{2}\quad$   &  $2$     &   $\textnormal{S}^g_{11}+\textnormal{S}^r_{31}+2\,\textnormal{S}^s_{41}-\frac{1}{3}\,(\textnormal{V}^r_{13}+2\,\textnormal{V}^s_{14})$ \\

            $\cellcolor{lred}\nicefrac{3}{2}$   &  \cellcolor{lred}$\quad 2\quad$     &    \cellcolor{lred} $\textnormal{S}^g_{12}-2\textnormal{S}^r_{41}-\frac{2}{3}(\textnormal{V}^s_{13}-2\textnormal{V}^r_{14})$  \\

            $\nicefrac{3}{2}$   &  $2$     &    $\textnormal{S}^g_{13}+2\,(\textnormal{S}^s_{43}-\textnormal{S}^s_{34})+\frac{2}{3}\,(\textnormal{V}^r_{11}+2\,\textnormal{V}^s_{12})$   \\

            $\nicefrac{3}{2}$   &  $2$     &    $\textnormal{S}^g_{14}-2\,(\textnormal{S}^r_{43}-\textnormal{S}^r_{34})+\frac{2}{3}\,(\textnormal{V}^s_{11}-2\,\textnormal{V}^r_{12})$  \\[0.1cm]

            \hline\hline\rule{0mm}{0.5cm}

            $\mathpzc{s}$           &  $\ell$  &   $\sqrt{5}\,\tau^{\sigma,1}_{3j}$   \\ [0.15cm]   \hline\hline
            $\quad\nicefrac{3}{2}\quad$   &  $2$     &   $\sqrt{5}(\textnormal{S}^g_{11}+3\,\textnormal{S}^r_{31}-\textnormal{V}^r_{13})$ \\

            $\nicefrac{3}{2}$   &  $\quad 3\quad$     &    $4\,\textnormal{S}^g_{12}+5\,(\textnormal{S}^r_{32}+\textnormal{S}^s_{42})+\textnormal{V}^r_{14}-\textnormal{V}^s_{13})$  \\

            $\cellcolor{lred}\nicefrac{3}{2}$   & \cellcolor{lred} $3$     &    \cellcolor{lred} $\textnormal{S}^g_{13}+5\,\textnormal{S}^r_{33}-\textnormal{V}^r_{11}$   \\

            $\nicefrac{3}{2}$   &  $3$     &    $\textnormal{S}^g_{14}+5\,\textnormal{S}^s_{44}-\textnormal{V}^s_{11}$  \\[0.1cm]

            \hline\hline\rule{0mm}{0.5cm}

            $\mathpzc{s}$           &  $\ell$  &   $\frac{1}{\sqrt{3}}\,\tau^{\sigma,1}_{4j}$   \\ [0.15cm]   \hline\hline
            $\quad\nicefrac{3}{2}\quad$   &  $3$     &   $\, \textnormal{S}^g_{11}+2\,(\textnormal{S}^s_{32}+\textnormal{S}^s_{41}+\textnormal{S}^r_{31})-\frac{2}{3}\,(\textnormal{V}^r_{13}+2\,\textnormal{V}^s_{14}) \,$ \\

            $\nicefrac{3}{2}$   &  $\quad 3\quad$     &    $\textnormal{S}^r_{32}-\textnormal{S}^s_{42}-\frac{1}{3}\,(\textnormal{V}^s_{13}+\textnormal{V}^r_{14})$  \\

            $\nicefrac{3}{2}$   &  $3$     &    $\textnormal{S}^g_{13}+\textnormal{S}^r_{33}+2\textnormal{S}^s_{43}+\frac{1}{3}\,(\textnormal{V}^r_{11}+2\,\textnormal{V}^s_{12})$   \\

            $\nicefrac{3}{2}$   &  $3$     &    $\textnormal{S}^g_{14}+\textnormal{S}^s_{44}+2\textnormal{S}^r_{34}+\frac{1}{3}\,(\textnormal{V}^s_{11}-2\,\textnormal{V}^r_{12})$  \\[0.1cm]
            \hline
            \end{tabular}
            \caption{Orthonormal Dirac basis $\mathrm{\tau}_{ij}^{\sigma,k}$  for $\mathpzc{s}=\nicefrac{3}{2}$ and $k=1$.
                    We omit Dirac and Lorentz indices as well as the label $\sigma=\pm$ for better readability.
                    The dominant covariants in each $(\mathpzc{s},\ell)$-sector which are plotted in Fig.~\ref{fig:amplitudes} are highlighted;
                    the relevant values of $\sigma$ and $k$ are given in the caption of Fig.~\ref{fig:amplitudes}.  }\label{table_basis3_2}
            \end{table}

            \begin{table}[t]
            \begin{tabular}{ | c | c || c  |  c |} \hline\rule{0mm}{0.5cm}

            $\mathpzc{s}$    &  $\ell$  &   $\frac{1}{\sqrt{3}}\,\tau^{\sigma,1}_{5j}$ &  $\tau^{\sigma,1}_{6j}$ \\[0.15cm] \hline\hline

            $\nicefrac{1}{2}$   &  $2$    &  $\textnormal{S}^r_{13}+\textnormal{S}^s_{14}$ & $\textnormal{V}^r_{13}+\textnormal{V}^s_{14}$ \\

            $\quad\nicefrac{1}{2}\quad$   &  $\quad 1\quad$    &  $\quad \textnormal{S}^r_{14}-\textnormal{S}^s_{13} \quad$ & $\quad \textnormal{V}^r_{14}-\textnormal{V}^s_{13} \quad$ \\

            $\cellcolor{lred}\nicefrac{1}{2}$   &  \cellcolor{lred}$1$    &  $\textnormal{S}^r_{11}$ & \cellcolor{lred}$\textnormal{V}^r_{11}$ \\

            $\nicefrac{1}{2}$   &  $1$    &  $\textnormal{S}^s_{11}$ & $\textnormal{V}^s_{11}$ \\[0.1cm]

            \hline\hline\rule{0mm}{0.5cm}

            $\mathpzc{s}$    &  $\ell$  &   $\tau^{\sigma,1}_{7j}$ &  $\sqrt{3}\,\tau^{\sigma,1}_{8j}$ \\[0.15cm] \hline\hline

            $\nicefrac{1}{2}$   &  $2$    &  $\textnormal{S}^r_{13}-\textnormal{S}^s_{14}$ & $\textnormal{V}^r_{13}-\textnormal{V}^s_{14}$ \\

            $\nicefrac{1}{2}$   &  $2$    &  $\textnormal{S}^s_{13}+\textnormal{S}^r_{14}$ & $\textnormal{V}^s_{13}+\textnormal{V}^r_{14}$ \\

            $\nicefrac{1}{2}$   &  $2$    &  $\textnormal{S}^r_{11}+2\,\textnormal{S}^s_{12}$ & $\textnormal{V}^r_{11}+2\,\textnormal{V}^s_{12}$ \\

            $\cellcolor{lred}\nicefrac{1}{2}$   &  \cellcolor{lred}$2$    &  \cellcolor{lred}$\textnormal{S}^s_{11}-2\,\textnormal{S}^r_{12}$ & $\textnormal{V}^s_{11}-2\,\textnormal{V}^r_{12}$ \\[0.1cm]
            \hline
            \end{tabular}
            \caption{Orthonormal Dirac basis $\mathrm{\tau}_{ij}^{\sigma,k}$
            for $\mathpzc{s}=\nicefrac{1}{2}$ and $k=1$. We omit Dirac/Lorentz indices and the label $\sigma$.
            The dominant covariants in each $(\mathpzc{s},\ell)$-sector are highlighted. }
            \label{table_basis1_2}
            \end{table}

    \renewcommand{\arraystretch}{1.4}

            To begin with,
            it is convenient to express the momenta
            $\{p,q,P\}$ that enter the basis elements through orthogonal unit vectors $\{r,s,\hat{P}\}$ which satisfy
            $r^2 = s^2 = \hat{P}^2 = 1$ and $r\cdot s = r\cdot \hat{P} = s\cdot\hat{P} = 0$.
            This is achieved by taking the component of $p$ transverse to $P$, and by
            projecting $q$ onto the direction transverse to $P$ and $r$:
               \begin{equation*}
                   r := \widehat{p_T} = \frac{\hat{p}-z_1 \hat{P}}{\sqrt{1-z_1^2}}\,, \quad
                   \widehat{q_T} = \frac{\hat{q}-z_2 \hat{P}}{\sqrt{1-z_2^2}}\,, \quad
                   s := \frac{\widehat{q_T}-z_0 \,r}{\sqrt{1-z_0^2}}\,
               \end{equation*}
            with the Lorentz-invariant momentum variables $z_1$, $z_2$ and $z_0$ from Eq.~\eqref{momentum-invariants}.
            In the baryon's rest frame and a suitable choice of coordinates,
            $r$, $s$ and $\hat{P}$ become the Euclidean unit vectors $e_3$, $e_2$ and $e_4$.

            Next, we define the basic Dirac structures
               \begin{equation}
                   \Gamma_{i=1\dots 4} = \{ \mathds{1}, \, \Slash{r} \Slash{s},\, \Slash{r},\,\Slash{s} \}
               \end{equation}
            which we use to construct a (still linearly dependent) basis for a rank-four Dirac tensor:
               \begin{eqnarray}\label{eq:basis_dirac}
               \left(\begin{array}{c}
               \textnormal{S}_{ij}^\sigma\\
               \textnormal{P}_{ij}^\sigma\\
               \textnormal{V}_{ij}^\sigma\\
               \textnormal{A}_{ij}^\sigma
               \end{array}\right):=
               \left(
               \begin{array}{c}
               \mathds{1}\otimes\mathds{1}\\
               \gamma^5\otimes\gamma^5\\
               \gamma^\mu_T\otimes\gamma^\mu_T\\
               \gamma^\mu_T\gamma^5\otimes\gamma^\mu_T\gamma^5
               \end{array}\right)\left(\Gamma_i\otimes \Gamma_j\right)
               \Omega^{\sigma}(\hat{P})\,.\qquad
               \end{eqnarray}
            Here, $\gamma^\mu_T$ is the $\gamma-$matrix transverse to $P$, and we used
            $\Omega^\pm(\hat{P})=\Lambda^\pm(\hat{P})\,\gamma^5\mathcal{C}\otimes \Lambda^+(\hat{P})$,
            where $\mathcal{C}=\gamma^4\gamma^2$ is  the charge conjugation-matrix and
            $\Lambda^\pm(\hat{P})=\left(\mathds{1}\pm\Slash{\hat{P}}\right)/2$
            are the positive- and negative-energy projectors.
            The tensor products are understood as
               \begin{equation}
               \begin{split}
                   (f\otimes g)_{\alpha\beta\gamma\delta} &= f_{\alpha\beta} \,g_{\gamma\delta}\,, \\
                   (f_1\otimes f_2)(g_1\otimes g_2) &= (f_1\,g_1)\otimes(f_2\, g_2)\,.
               \end{split}
               \end{equation}

            Denoting the elements in Eq.~\eqref{eq:basis_dirac} generically by $M_{ij}^\sigma$, with
            $M\in\{\textnormal{S},\textnormal{P},\textnormal{V},\textnormal{A}\}$ and $i,j=1\dots 4$,
            we can construct the building blocks of a
            Dirac-Lorentz basis for the $\Delta-$baryon in the following way:
               \begin{eqnarray}\label{eq:basis_mixed}
               \left(\begin{array}{c}
               \left[M^g_{ij}\right]^{\sigma,\nu}\\
               \left[M^r_{ij}\right]^{\sigma,\nu}\\
               \left[M^s_{ij}\right]^{\sigma,\nu}
               \end{array}\right):=
               \left(
               \begin{array}{r}
               \gamma^\mu_T\,\gamma^5\otimes\mathds{1}\\
               r^\mu\gamma^5\otimes\mathds{1}\\
               s^\mu\gamma^5\otimes\mathds{1}
               \end{array}\right)(M_{ij}^\sigma)\,(\mathds{1}\otimes\mathbb{P}^{\mu\nu})\,,\quad
               \quad
               \end{eqnarray}
            where $\mathbb{P}^{\mu\nu}$ is the Rarita-\textnormal{S}chwinger projector for
            positive-energy particles:
            \begin{equation}
             \mathbb{P}_+^{\mu\nu}(\hat{P})=\Lambda^+(\hat{P})
             \left(T_P^{\mu\nu}-\frac{1}{3}\gamma^\mu_T\gamma^\nu_T\right)\,.
            \end{equation}
            The set (\ref{eq:basis_mixed})
            contains 384 elements, but one can show that only 128 of them are
            linearly independent. We found the following choice of linearly independent
            elements convenient:
               \begin{equation}\label{eq:LI_set}
               \begin{array}{c}
               \left\{
                     \begin{array}{c}
                      \textnormal{S}^r_{1j}\,,\textnormal{S}^s_{1j}\\
                      \textnormal{P}^r_{1j}\,,\textnormal{P}^s_{1j}\\
                      \textnormal{V}^r_{1j}\,,\textnormal{V}^s_{1j}\\
                      \textnormal{A}^r_{1j}\,,\textnormal{A}^s_{1j}
                     \end{array}
               \right\},\quad
               \left\{
                     \begin{array}{c}
                      \textnormal{S}^r_{43}\,,\textnormal{S}^r_{41}\\
                      \textnormal{S}^s_{32}\,,\textnormal{S}^s_{34}\\
                      \textnormal{P}^r_{43}\,,\textnormal{P}^r_{41}\\
                      \textnormal{P}^s_{32}\,,\textnormal{P}^s_{34}
                     \end{array}
               \right\},\\[1.1cm]
               \left\{
                     \begin{array}{c}
                      \textnormal{S}^g_{1j}\\
                      \textnormal{P}^g_{1j}
                     \end{array}
               \right\},\quad
               \left\{
                     \begin{array}{c}
                      \textnormal{S}^r_{3j}\,,\textnormal{S}^s_{4j}\\
                      \textnormal{P}^r_{3j}\,,\textnormal{P}^s_{4j}
                     \end{array}
               \right\},
               \end{array}
               \end{equation}
            with $j=1\dots4$, and we have omitted the index $\sigma$ for better readability.
            The set~\eqref{eq:LI_set} includes the element
               \begin{equation}
               \begin{split}
                    [\textnormal{S}^g_{11}]^{\sigma,\nu} &= (\gamma^\mu_T\,\gamma^5\otimes\mathds{1}) \,\Omega^\sigma(\hat{P})\,(\mathds{1}\otimes\mathbb{P}^{\mu\nu}) = \\
                                                   &= \Lambda^\sigma(\hat{P})\,\gamma^\mu_T \,\mathcal{C} \otimes \mathbb{P}^{\mu\nu}
               \end{split}
               \end{equation}
            which is familiar from the quark-diquark model and represents the dominant $s-$wave structure in the $\Delta-$baryon amplitude.

            Let us now turn to the partial-wave analysis of these basis elements.
            In the baryon rest frame, the quark total spin and
            relative angular momentum operators read
               \begin{equation}\label{eq:SL_operators}
               \begin{array}{l}
               S^2=\frac{9}{4}\left(\mathds{1}\otimes\mathds{1}
               \otimes\mathds{1}\right)+\frac{1}{2}\left(\sigma^{\mu\nu}_T
               \otimes\sigma^{\mu\nu}_T\otimes\mathds{1}+\textnormal{perm.}\right),\quad\\
               L^2=L^2_{(p)}+L^2_{(q)}+2\,L_{(p)}\cdot L_{(q)}\,,
               \end{array}
               \end{equation}
            where $\sigma_T^{\mu\nu} = -\frac{i}{2}\,[\gamma_T^\mu,\gamma_T^\nu]$ and
               \begin{equation}\label{eq:L_operators}
               \begin{array}{rcl}
               L^2_{(p)}&=&2\,\textbf{p}\cdot\nabla_\textbf{p}
               +p^i(\textbf{p}\cdot\nabla_\textbf{p})\nabla^i_\textbf{p}
               -\textbf{p}^2\Delta_\textbf{p}\,,\quad\\
               L^2_{(q)}&=&2\,\textbf{q}\cdot\nabla_\textbf{q}
               +q^i(\textbf{q}\cdot\nabla_\textbf{q})\nabla^i_\textbf{q}
               -\textbf{q}^2\Delta_\textbf{q}\,,\quad \\
               L_{(p)}\cdot L_{(q)}&=&
               p^i(\textbf{q}\cdot\nabla_\textbf{p})\nabla^i_\textbf{q}
               -(\textbf{p}\cdot\textbf{q})(\nabla_\textbf{p}\cdot\nabla_\textbf{q})\,,
               \end{array}
               \end{equation}
            and $\textbf{p}$, $\textbf{q}$ are the spatial parts of $p_T$ and $q_T$,
            respectively.

            It is useful to realize that the basis elements
            containing $\{\textnormal{S},\textnormal{V}\}$ and
            $\{\textnormal{P},\textnormal{A}\}$, which differ by a factor
            $\gamma^5\otimes\gamma^5$, and those with a different value for the
            index $\sigma=\pm$, do not mix under the action of
            $\textnormal{S}^2$ or $L^2$ and thus can be analyzed independently.
            Moreover, from Eqs.~(\ref{eq:SL_operators}) and
            (\ref{eq:L_operators}) one infers that the set (\ref{eq:LI_set})
            can be further subdivided into four subsets which, due to their different
            momentum dependence, again do not mix under
            $\textnormal{S}^2$ or $L^2$:
               \begin{equation*}
               \begin{array}{ll}
               1,\,r^2,\,s^2,\,r^2s^2:&     \textnormal{S}^g_{11}, \,
                                            \textnormal{S}^r_{13},\,
                                            \textnormal{S}^s_{14},\,
                                            \textnormal{S}^r_{31},\,
                                            \textnormal{S}^s_{41},\,
                                            \textnormal{S}^s_{32},\,
                                            \textnormal{V}^r_{13},\,
                                            \textnormal{V}^s_{14}  \\
               rs,\,r^3s,\,rs^3:&           \textnormal{S}^g_{12},\,
                                            \textnormal{S}^r_{14},\,
                                            \textnormal{S}^s_{13},\,
                                            \textnormal{S}^s_{42},\,
                                            \textnormal{S}^r_{41},\,
                                            \textnormal{S}^r_{32},\,
                                            \textnormal{V}^r_{14},\,
                                            \textnormal{V}^s_{13}\\
               r,\,rs^2,\,r^3: &            \textnormal{S}^g_{13},\,
                                            \textnormal{S}^r_{11},\,
                                            \textnormal{S}^s_{12},\,
                                            \textnormal{S}^r_{33},\,
                                            \textnormal{S}^s_{34},\,
                                            \textnormal{S}^s_{43},\,
                                            \textnormal{V}^r_{11},\,
                                            \textnormal{V}^s_{12}\\
               s,\,r^2s,\,s^3: &            \textnormal{S}^g_{14},\,
                                            \textnormal{S}^r_{12},\,
                                            \textnormal{S}^s_{11},\,
                                            \textnormal{S}^s_{44},\,
                                            \textnormal{S}^r_{34},\,
                                            \textnormal{S}^r_{43},\,
                                            \textnormal{V}^r_{12},\,
                                            \textnormal{V}^s_{11}
               \end{array}
               \end{equation*}
            Here, the left column symbolically indicates the different momentum  dependencies
            of the basis elements in powers of  $r$ and $s$.
            The partial-wave analysis is now considerably simplified since we only need
            to look for eigenfunctions of $S^2$ and $L^2$ within the above subsets.

            The operator $S^2$ is independent of the momentum content of the basis
            elements. Therefore it is sufficient to find the eigenstates at fixed values
            for the unit vectors $r$ and $s$. At this point the problem can
            be easily implemented and solved using a symbolic programming language.
            Similarly, the orbital angular-momentum decomposition can be done
            straightforwardly on a computer, but it is nevertheless instructive to study some simple
            examples. The $\ell=0$ elements can be found immediately: they are the
            momentum-independent elements in (\ref{eq:LI_set}),
            \textit{i.e.},
            $[\textnormal{S}^g_{11}]^\sigma$ and $[\textnormal{P}^g_{11}]^\sigma$.
            The remaining basis elements can be written as
            contractions of
            \begin{equation}
             r^\alpha,\,\, s^\alpha,\,\, r^\alpha s^\beta,\,\, r^\alpha r^\beta,\,\,
             s^\alpha s^\beta,\,\, r^\alpha r^\beta s^\delta,\,\,\dots
            \end{equation}
            with appropriate momentum-independent Dirac-Lorentz tensors.
            For the $\ell=1$ elements it is sufficient to consider the first three elements
            in the list above. Applying the orbital angular-momentum operator yields
            \begin{equation}
            \begin{array}{c}
              L^2 r^\alpha=2\,r^\alpha\,,\\
              L^2 s^\alpha=2\,s^\alpha\,,\\
              L^2 r^\alpha s^\beta=4\,r^\alpha s^\beta+2\,s^\alpha r^\beta\,,
            \end{array}
            \end{equation}
            and thus the $\ell=1$ eigenfunctions are given by:
            \begin{equation}
              r^\alpha,\quad s^\alpha, \quad r^\alpha s^\beta-s^\alpha r^\beta\,.
            \end{equation}
            Their contraction with the corresponding Dirac-Lorentz structures
            yields the $\ell=1$ eigentensors at the level of the basis elements.
            For other $\ell$ values the calculation proceeds along the same lines but
            is more involved.

            The above analysis focused on the subset $\{\textnormal{S},\textnormal{V}\}$,
            which we denote by the index $k=1$. The final result of
            the partial-wave decomposition for $k=1$ is presented in Tables~\ref{table_basis3_2}
            and \ref{table_basis1_2}.
            Analogous results hold for the set $\{\textnormal{P},\textnormal{A}\}$, denoted by k=2:
            the respective basis elements are obtained
            by exchanging $\textnormal{S}\rightarrow\textnormal{P}$,
            $\textnormal{V}\rightarrow\textnormal{A}$ and adding an extra minus sign
            to the elements $\textnormal{P}^g_{1j}$.
            For the compilation in Tables~(\ref{table_basis3_2}--\ref{table_basis1_2}),
            the original label $i=1 \dots 128$ that appears in Eq.~\eqref{eq:basis_decomposition} was
            dissolved into the indices $i=1\dots 8$, $j=1\dots 4$,
            $k=1,2$, and $\sigma=\pm$.

            The resulting basis satisfies the following orthonormality relation:
               \begin{equation}\label{basis-orthogonality}
               \begin{split}
                \textstyle\frac{1}{8}\,\textnormal{Tr}\left\{\bar{\tau}^{\sigma,k}_{ij}
                \tau^{\sigma',k'}_{i'j'}\right\} &= \textstyle\frac{1}{8}
               \big[\bar{\tau}^{\sigma,k}_{ij}\big]_{\beta\alpha\delta\gamma}^\mu
               \big[\tau^{\sigma',k'}_{i'j'}\big]_{\alpha\beta\gamma\delta}^\mu= \\
               &=\delta_{ii'}\delta_{jj'}\delta_{kk'}\delta_{\sigma\sigma'}\,,
               \end{split}
               \end{equation}
            where the conjugation of the basis elements is defined as
              \begin{equation}
              \begin{split}
               \bar{\tau}^\mu_{\alpha\beta\gamma\delta}(p,q,P) &= \\ -C_{\alpha \alpha'} \,C_{\gamma \gamma'} \,&
               \tau^\mu_{\beta'\alpha'\delta'\gamma'}(-p,-q,-P)\, C_{\beta'\beta}^T \,C_{\delta'\delta}^T\,.
              \end{split}
              \end{equation}

                  \begin{figure}[t]
                  \begin{center}
                  \includegraphics[scale=0.95]{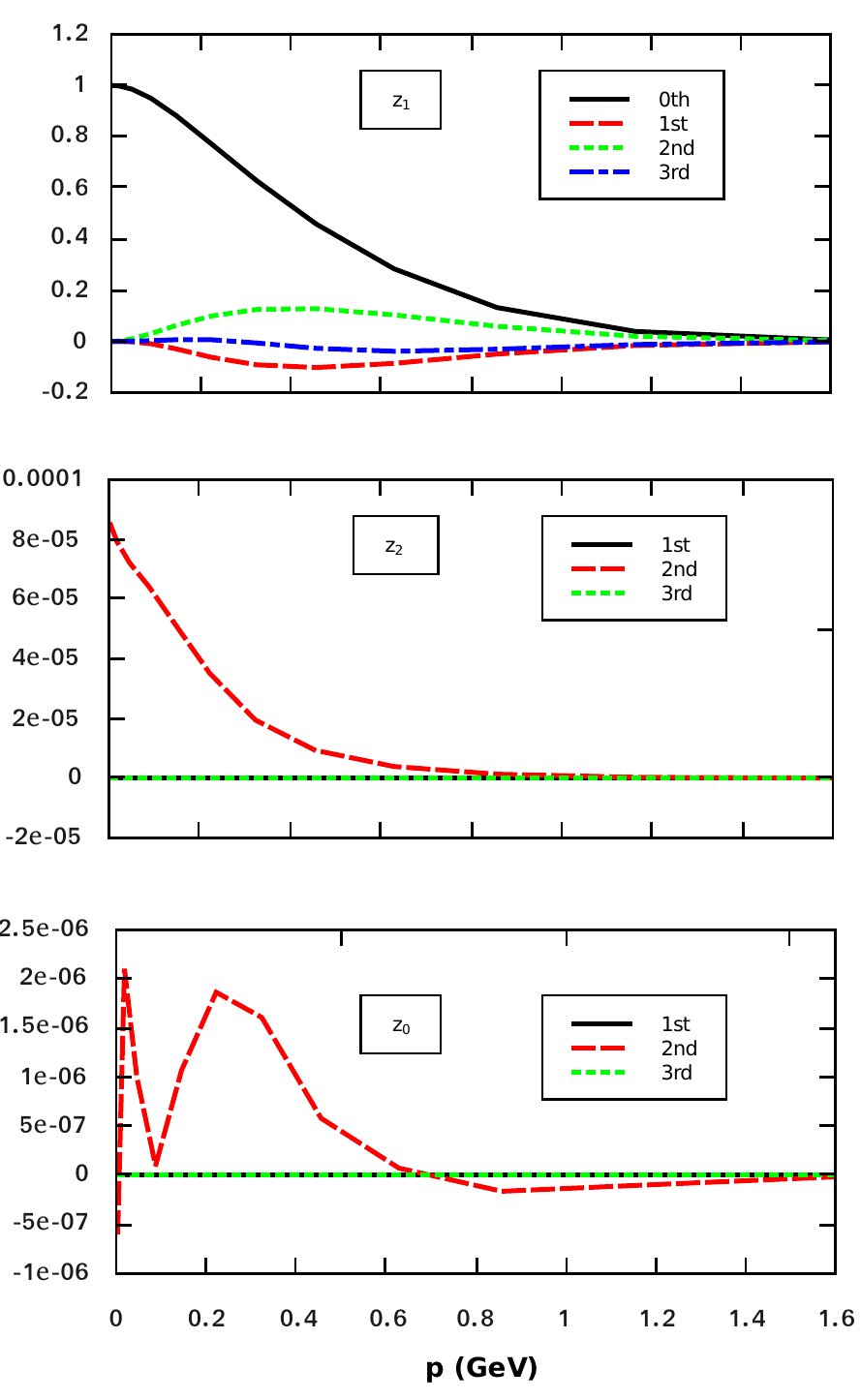}
                  \caption{First four Chebyshev moments in the variables
                  $z_0$, $z_1$ and $z_2$ for the dominant amplitude $[S^g_{11}]^+$ at $q^2=0$.
                  Note that the zeroth order is the same for all three figures; in the lower two
                  panels it is not shown for clarity.}
                  \label{fig:chebys_moments}
                  \end{center}
                  \end{figure}

\section{Numerical details}\label{sec:Numerics}

              The numerical techniques used in this work are an extension of those explained
              in Ref.\ \cite{Eichmann:2011vu} towards the case of the $\Delta$-baryon. We
              summarize the main ideas in this appendix.

              The Faddeev amplitude is the product of color, flavor and spin parts. As
              required by the Pauli principle, that product must be antisymmetric under the
              exchange of any two of the three quarks. Since a baryon is a color singlet, the
              color part of the Faddeev amplitude is always antisymmetric, and thus the
              product of flavor and spin parts must be symmetric. This symmetry requirement
              allows to relate the three diagrams in Fig.\ \ref{fig:faddeev-eq}.

              Compared to the nucleon, the situation is simpler for the $\Delta$- and
              $\Omega$-baryons. Since the $\Delta$ carries
              isospin $\nicefrac{3}{2}$, its flavor part is completely symmetric and
              hence also the spin part must be symmetric. The $\Omega$, on the other hand, is a pure s-quark state
              and thus automatically symmetric in flavor space. Using this, and starting
              from Eq.\ (\ref{eq:faddeev_eq}), it is not difficult to show that the first
              two terms in the equation can be reexpressed in terms of the third one,
              evaluated at different momenta:
              \begin{equation}\label{eq:faddev_eq_new}
              \begin{split}
               \Psi_{\alpha\beta\gamma\delta}\,^\mu(p,q,P)&=
               \Psi_{\alpha\beta\gamma\delta}^{(3)}\,^\mu(p,q,P)\\
              &+\Psi_{\beta\gamma\alpha\delta}^{(3)}\,^\mu(p',q',P)\\
              &+\Psi_{\gamma\alpha\beta\delta}^{(3)}\,^\mu(p'',q'',P) \,,
              \end{split}
              \end{equation}
              where the permuted relative momenta are given by
 \renewcommand{\arraystretch}{1.3}
             \begin{equation}\label{perm-momenta}
                 \begin{array}{rl}
                     p' &= -q-\frac{p}{2}\,, \\
                     p'' &= q-\frac{p}{2}\,,
                 \end{array}\qquad
                 \begin{array}{rl}
                     q' &= -\frac{q}{2} + \frac{3p}{4}\,,                      \\
                     q'' &= -\frac{q}{2} - \frac{3p}{4}\,.
                 \end{array}
             \end{equation}
 \renewcommand{\arraystretch}{1.2}

              In terms of the amplitude decomposition (\ref{eq:basis_decomposition}),
              the Faddeev equation is written as
              \begin{equation}
               f_i(p,q,P)=\sum_{a=1}^3 f_i^{(a)}(p,q,P)\,,
              \end{equation}
              where we used the notation
              \begin{equation}\label{eq:faddev_eq_newb}
              \begin{split}
               f_i^{(a)}(p,q,P)&=\int_k\mathcal{K}_{ij}^{(a)}(p,q,P;k) \,\phi_j^{(a)}(p^{(a)},q^{(a)},P)\,, \\
               \phi_i^{(a)}(p,q,P)&=\mathcal{G}_{ij}^{(a)}(p,q,P)\,f_j(p,q,P)\,,
              \end{split}
              \end{equation}
              with $p^{(a)}$ and $q^{(a)}$ being the internal relative momenta from Eq.~\eqref{internal-relative-momenta}.
              The functions $\phi_i^{(a)}$ are the coefficients
              of the so-called wave functions $\Phi^{(a)}=S(p_b)S(p_c)\Psi$ in the basis expansion~\eqref{eq:basis_decomposition},
              where $\{a,b,c\}$ is a symmetric permutation of $\{1,2,3\}$,
              and $\mathcal{K}_{ij}$ and $\mathcal{G}_{ij}$ are the matrix
              representations of the kernel and propagator operators in that basis.
              Eq.~(\ref{eq:faddev_eq_new}) can be written in terms of the coefficients
              $f_i(p,q,P)$ as
              \begin{equation}\label{eq:faddev_eq_new2}
              \begin{split}
                f_i(p,q,P) &=f_i^{(3)}(p,q,P)  \\
                           &+H'_{ij} \, f_j^{(3)}(p',q',P) \\
                           &+H''_{ij} \, f_j^{(3)}(p'',q'',P)\, ,
              \end{split}
              \end{equation}
              where the matrices $H'$ and $H''$ are defined by
              \begin{equation}
               \begin{split}
                H'_{ij}(p,q,P)&=\textstyle\frac{1}{8}\,\bar{\tau}^i_{\beta\alpha\delta\gamma}\,^\mu(p,q,P)\,
                \tau^j_{\beta\gamma\alpha\delta}\,^\mu(p',q',P)\,,\\
                H''_{ij}(p,q,P)&=\textstyle\frac{1}{8}\,\bar{\tau}^i_{\beta\alpha\delta\gamma}\,^\mu(p,q,P)\,
                \tau^j_{\gamma\alpha\beta\delta}\,^\mu(p'',q'',P)\,.
               \end{split}
              \end{equation}
              In rewriting the equation we have used the orthogonality~\eqref{basis-orthogonality} of the basis.

              The algorithm for solving the Faddeev equation begins by introducing a
              fictitious eigenvalue $\lambda(M)$  in front of the integral in
              (\ref{eq:faddev_eq_newb}). The integral equation is then solved by iteration:
              start with a guess for the bound-state mass $M$ and the initial amplitudes
              $f_i$; compute the integral~\eqref{eq:faddev_eq_newb} for the case $a=3$;
              apply the permutation~\eqref{eq:faddev_eq_new2}; and repeat until the eigenvalue $\lambda(M)$ has
              converged. If the converged value is $\lambda(M)=1$, then $M$ is the correct
              mass for the bound state; otherwise one has to change the guess for $M$ and
              repeat the procedure.

              For the presented solutions of the Faddeev equation we expanded  the angular
              dependence of the Faddeev amplitudes in terms of Chebyshev  polynomials. This
              is convenient because only a small number of Chebyshev moments  are needed, as
              shown in Fig.\ \ref{fig:chebys_moments}.

              Solving the Faddeev equation in the form (\ref{eq:faddev_eq_newb}--\ref{eq:faddev_eq_new2}) is a great
              simplification with respect to the original formulation (\ref{eq:faddeev_eq})
              since the kernel matrix $\mathcal{K}^{(3)}$ is comparatively simpler than the
              other two permuted kernels. Note that in this case the external and internal
              relative momenta $p$ and $p^{(3)}$ are the same, cf.~Eq.~\eqref{internal-relative-momenta}. If one uses a moderate number
              of integration points, and taking into account the fact that the kernel
              matrices are typically sparse, then the kernel matrix can be computed and
              stored in advance which notably reduces the  computing time. However, if the
              number of integration points is increased, memory limitations become an issue.
              For the presented results
              we have used the following number of integration points: $20$ for
              $\{p^2,q^2,k^2\}$, $8$ for $z_0$ and $4$ for $\{z_1,z_2,z,y\}$. For this
              configuration, and without storing the kernel matrix in advance (although it
              would be feasible), each iteration requires approximately
              four CPU-hours on a
              16-node 2.66~GHz cluster, and convergence for $\lambda(M)$ is usually reached within 15-20
              iterations.

\end{appendix}

\bibliographystyle{apsrev4-1-mod}

\bibliography{lit-delta}

\begin{thebibliography}{47}%
\makeatletter
\providecommand \@ifxundefined [1]{%
 \@ifx{#1\undefined}
}%
\providecommand \@ifnum [1]{%
 \ifnum #1\expandafter \@firstoftwo
 \else \expandafter \@secondoftwo
 \fi
}%
\providecommand \@ifx [1]{%
 \ifx #1\expandafter \@firstoftwo
 \else \expandafter \@secondoftwo
 \fi
}%
\providecommand \natexlab [1]{#1}%
\providecommand \enquote  [1]{``#1''}%
\providecommand \bibnamefont  [1]{#1}%
\providecommand \bibfnamefont [1]{#1}%
\providecommand \citenamefont [1]{#1}%
\providecommand \href@noop [0]{\@secondoftwo}%
\providecommand \href [0]{\begingroup \@sanitize@url \@href}%
\providecommand \@href[1]{\@@startlink{#1}\@@href}%
\providecommand \@@href[1]{\endgroup#1\@@endlink}%
\providecommand \@sanitize@url [0]{\catcode `\\12\catcode `\$12\catcode
  `\&12\catcode `\#12\catcode `\^12\catcode `\_12\catcode `\%12\relax}%
\providecommand \@@startlink[1]{}%
\providecommand \@@endlink[0]{}%
\providecommand \url  [0]{\begingroup\@sanitize@url \@url }%
\providecommand \@url [1]{\endgroup\@href {#1}{\urlprefix }}%
\providecommand \urlprefix  [0]{URL }%
\providecommand \Eprint [0]{\href }%
\providecommand \doibase [0]{http://dx.doi.org/}%
\providecommand \selectlanguage [0]{\@gobble}%
\providecommand \bibinfo  [0]{\@secondoftwo}%
\providecommand \bibfield  [0]{\@secondoftwo}%
\providecommand \translation [1]{[#1]}%
\providecommand \BibitemOpen [0]{}%
\providecommand \bibitemStop [0]{}%
\providecommand \bibitemNoStop [0]{.\EOS\space}%
\providecommand \EOS [0]{\spacefactor3000\relax}%
\providecommand \BibitemShut  [1]{\csname bibitem#1\endcsname}%
\let\auto@bib@innerbib\@empty
\bibitem [{\citenamefont {Salpeter}\ and\ \citenamefont
  {Bethe}(1951)}]{Salpeter:1951sz}%
  \BibitemOpen
  \bibfield  {author} {\bibinfo {author} {\bibfnamefont {E.~E.}\ \bibnamefont
  {Salpeter}}\ and\ \bibinfo {author} {\bibfnamefont {H.~A.}\ \bibnamefont
  {Bethe}},\ }\href {\doibase 10.1103/PhysRev.84.1232} {\bibfield  {journal}
  {\bibinfo  {journal} {Phys. Rev.}\ }\textbf {\bibinfo {volume} {84}},\
  \bibinfo {pages} {1232} (\bibinfo {year} {1951})}\BibitemShut {NoStop}%
\bibitem [{\citenamefont {Faddeev}(1961)}]{Faddeev:1960su}%
  \BibitemOpen
  \bibfield  {author} {\bibinfo {author} {\bibfnamefont {L.~D.}\ \bibnamefont
  {Faddeev}},\ }\href@noop {} {\bibfield  {journal} {\bibinfo  {journal} {Sov.
  Phys. JETP}\ }\textbf {\bibinfo {volume} {12}},\ \bibinfo {pages} {1014}
  (\bibinfo {year} {1961})},\ \bibinfo {note} {[Zh. Eksp. Teor. Fiz. {\bf 39},
  1014 (1960)]}\BibitemShut {NoStop}%
\bibitem [{\citenamefont {Taylor}(1966)}]{Taylor:1966zza}%
  \BibitemOpen
  \bibfield  {author} {\bibinfo {author} {\bibfnamefont {J.~G.}\ \bibnamefont
  {Taylor}},\ }\href {\doibase 10.1103/PhysRev.150.1321} {\bibfield  {journal}
  {\bibinfo  {journal} {Phys. Rev.}\ }\textbf {\bibinfo {volume} {150}},\
  \bibinfo {pages} {1321} (\bibinfo {year} {1966})}\BibitemShut {NoStop}%
\bibitem [{\citenamefont {Boehm}\ and\ \citenamefont
  {Meyer}(1979)}]{Boehm:1976ya}%
  \BibitemOpen
  \bibfield  {author} {\bibinfo {author} {\bibfnamefont {M.}~\bibnamefont
  {Boehm}}\ and\ \bibinfo {author} {\bibfnamefont {R.~F.}\ \bibnamefont
  {Meyer}},\ }\href {\doibase 10.1016/0003-4916(79)90395-6} {\bibfield
  {journal} {\bibinfo  {journal} {Annals Phys.}\ }\textbf {\bibinfo {volume}
  {120}},\ \bibinfo {pages} {360} (\bibinfo {year} {1979})}\BibitemShut
  {NoStop}%
\bibitem [{\citenamefont {Loring}\ \emph {et~al.}(2001)\citenamefont {Loring},
  \citenamefont {Kretzschmar}, \citenamefont {Metsch},\ and\ \citenamefont
  {Petry}}]{Loring:2001kv}%
  \BibitemOpen
  \bibfield  {author} {\bibinfo {author} {\bibfnamefont {U.}~\bibnamefont
  {Loring}}, \bibinfo {author} {\bibfnamefont {K.}~\bibnamefont {Kretzschmar}},
  \bibinfo {author} {\bibfnamefont {B.~C.}\ \bibnamefont {Metsch}}, \ and\
  \bibinfo {author} {\bibfnamefont {H.~R.}\ \bibnamefont {Petry}},\ }\href
  {\doibase 10.1007/s100500170117} {\bibfield  {journal} {\bibinfo  {journal}
  {Eur. Phys. J.}\ }\textbf {\bibinfo {volume} {A10}},\ \bibinfo {pages} {309}
  (\bibinfo {year} {2001})}\BibitemShut {NoStop}%
\bibitem [{\citenamefont {Alkofer}\ and\ \citenamefont {von
  Smekal}(2001)}]{Alkofer:2000wg}%
  \BibitemOpen
  \bibfield  {author} {\bibinfo {author} {\bibfnamefont {R.}~\bibnamefont
  {Alkofer}}\ and\ \bibinfo {author} {\bibfnamefont {L.}~\bibnamefont {von
  Smekal}},\ }\href {\doibase 10.1016/S0370-1573(01)00010-2} {\bibfield
  {journal} {\bibinfo  {journal} {Phys. Rept.}\ }\textbf {\bibinfo {volume}
  {353}},\ \bibinfo {pages} {281} (\bibinfo {year} {2001})}\BibitemShut
  {NoStop}%
\bibitem [{\citenamefont {Fischer}(2006)}]{Fischer:2006ub}%
  \BibitemOpen
  \bibfield  {author} {\bibinfo {author} {\bibfnamefont {C.~S.}\ \bibnamefont
  {Fischer}},\ }\href {\doibase 10.1088/0954-3899/32/8/R02} {\bibfield
  {journal} {\bibinfo  {journal} {J. Phys.}\ }\textbf {\bibinfo {volume}
  {G32}},\ \bibinfo {pages} {R253} (\bibinfo {year} {2006})}\BibitemShut
  {NoStop}%
\bibitem [{\citenamefont {Chang}\ \emph
  {et~al.}(2011{\natexlab{a}})\citenamefont {Chang}, \citenamefont {Roberts},\
  and\ \citenamefont {Tandy}}]{Chang:2011vu}%
  \BibitemOpen
  \bibfield  {author} {\bibinfo {author} {\bibfnamefont {L.}~\bibnamefont
  {Chang}}, \bibinfo {author} {\bibfnamefont {C.~D.}\ \bibnamefont {Roberts}},
  \ and\ \bibinfo {author} {\bibfnamefont {P.~C.}\ \bibnamefont {Tandy}},\
  }\href@noop {} {\ }\Eprint {http://arxiv.org/abs/1107.4003} {1107.4003
  [nucl-th]} \BibitemShut {NoStop}%
\bibitem [{\citenamefont {Maris}\ and\ \citenamefont
  {Tandy}(2006)}]{Maris:2005tt}%
  \BibitemOpen
  \bibfield  {author} {\bibinfo {author} {\bibfnamefont {P.}~\bibnamefont
  {Maris}}\ and\ \bibinfo {author} {\bibfnamefont {P.~C.}\ \bibnamefont
  {Tandy}},\ }\href {\doibase 10.1016/j.nuclphysbps.2006.08.012} {\bibfield
  {journal} {\bibinfo  {journal} {Nucl. Phys. Proc. Suppl.}\ }\textbf {\bibinfo
  {volume} {161}},\ \bibinfo {pages} {136} (\bibinfo {year}
  {2006})}\BibitemShut {NoStop}%
\bibitem [{\citenamefont {Maris}(2007)}]{Maris:2006ea}%
  \BibitemOpen
  \bibfield  {author} {\bibinfo {author} {\bibfnamefont {P.}~\bibnamefont
  {Maris}},\ }\href {\doibase 10.1063/1.2714348} {\bibfield  {journal}
  {\bibinfo  {journal} {AIP Conf. Proc.}\ }\textbf {\bibinfo {volume} {892}},\
  \bibinfo {pages} {65} (\bibinfo {year} {2007})}\BibitemShut {NoStop}%
\bibitem [{\citenamefont {Krassnigg}(2009)}]{Krassnigg:2009zh}%
  \BibitemOpen
  \bibfield  {author} {\bibinfo {author} {\bibfnamefont {A.}~\bibnamefont
  {Krassnigg}},\ }\href {\doibase 10.1103/PhysRevD.80.114010} {\bibfield
  {journal} {\bibinfo  {journal} {Phys. Rev.}\ }\textbf {\bibinfo {volume}
  {D80}},\ \bibinfo {pages} {114010} (\bibinfo {year} {2009})}\BibitemShut
  {NoStop}%
\bibitem [{\citenamefont {Carimalo}(1993)}]{Carimalo:1992ia}%
  \BibitemOpen
  \bibfield  {author} {\bibinfo {author} {\bibfnamefont {C.}~\bibnamefont
  {Carimalo}},\ }\href {\doibase 10.1063/1.530334} {\bibfield  {journal}
  {\bibinfo  {journal} {J. Math. Phys.}\ }\textbf {\bibinfo {volume} {34}},\
  \bibinfo {pages} {4930} (\bibinfo {year} {1993})}\BibitemShut {NoStop}%
\bibitem [{\citenamefont {Eichmann}\ \emph
  {et~al.}(2010{\natexlab{a}})\citenamefont {Eichmann}, \citenamefont
  {Alkofer}, \citenamefont {Krassnigg},\ and\ \citenamefont
  {Nicmorus}}]{Eichmann:2009en}%
  \BibitemOpen
  \bibfield  {author} {\bibinfo {author} {\bibfnamefont {G.}~\bibnamefont
  {Eichmann}}, \bibinfo {author} {\bibfnamefont {R.}~\bibnamefont {Alkofer}},
  \bibinfo {author} {\bibfnamefont {A.}~\bibnamefont {Krassnigg}}, \ and\
  \bibinfo {author} {\bibfnamefont {D.}~\bibnamefont {Nicmorus}},\ }\href
  {\doibase 10.1051/epjconf/20100303028} {\bibfield  {journal} {\bibinfo
  {journal} {EPJ Web Conf.}\ }\textbf {\bibinfo {volume} {3}},\ \bibinfo
  {pages} {03028} (\bibinfo {year} {2010}{\natexlab{a}})}\BibitemShut {NoStop}%
\bibitem [{\citenamefont {Hellstern}\ \emph {et~al.}(1997)\citenamefont
  {Hellstern}, \citenamefont {Alkofer}, \citenamefont {Oettel},\ and\
  \citenamefont {Reinhardt}}]{Hellstern:1997pg}%
  \BibitemOpen
  \bibfield  {author} {\bibinfo {author} {\bibfnamefont {G.}~\bibnamefont
  {Hellstern}}, \bibinfo {author} {\bibfnamefont {R.}~\bibnamefont {Alkofer}},
  \bibinfo {author} {\bibfnamefont {M.}~\bibnamefont {Oettel}}, \ and\ \bibinfo
  {author} {\bibfnamefont {H.}~\bibnamefont {Reinhardt}},\ }\href {\doibase
  10.1016/S0375-9474(97)00514-9} {\bibfield  {journal} {\bibinfo  {journal}
  {Nucl. Phys.}\ }\textbf {\bibinfo {volume} {A627}},\ \bibinfo {pages} {679}
  (\bibinfo {year} {1997})}\BibitemShut {NoStop}%
\bibitem [{\citenamefont {Oettel}\ \emph {et~al.}(1998)\citenamefont {Oettel},
  \citenamefont {Hellstern}, \citenamefont {Alkofer},\ and\ \citenamefont
  {Reinhardt}}]{Oettel:1998bk}%
  \BibitemOpen
  \bibfield  {author} {\bibinfo {author} {\bibfnamefont {M.}~\bibnamefont
  {Oettel}}, \bibinfo {author} {\bibfnamefont {G.}~\bibnamefont {Hellstern}},
  \bibinfo {author} {\bibfnamefont {R.}~\bibnamefont {Alkofer}}, \ and\
  \bibinfo {author} {\bibfnamefont {H.}~\bibnamefont {Reinhardt}},\ }\href
  {\doibase 10.1103/PhysRevC.58.2459} {\bibfield  {journal} {\bibinfo
  {journal} {Phys. Rev.}\ }\textbf {\bibinfo {volume} {C58}},\ \bibinfo {pages}
  {2459} (\bibinfo {year} {1998})}\BibitemShut {NoStop}%
\bibitem [{\citenamefont {Bloch}\ \emph {et~al.}(1999)\citenamefont {Bloch},
  \citenamefont {Roberts}, \citenamefont {Schmidt}, \citenamefont {Bender},\
  and\ \citenamefont {Frank}}]{Bloch:1999ke}%
  \BibitemOpen
  \bibfield  {author} {\bibinfo {author} {\bibfnamefont {J.~C.~R.}\
  \bibnamefont {Bloch}}, \bibinfo {author} {\bibfnamefont {C.~D.}\ \bibnamefont
  {Roberts}}, \bibinfo {author} {\bibfnamefont {S.~M.}\ \bibnamefont
  {Schmidt}}, \bibinfo {author} {\bibfnamefont {A.}~\bibnamefont {Bender}}, \
  and\ \bibinfo {author} {\bibfnamefont {M.~R.}\ \bibnamefont {Frank}},\ }\href
  {\doibase 10.1103/PhysRevC.60.062201} {\bibfield  {journal} {\bibinfo
  {journal} {Phys. Rev.}\ }\textbf {\bibinfo {volume} {C60}},\ \bibinfo {pages}
  {062201} (\bibinfo {year} {1999})}\BibitemShut {NoStop}%
\bibitem [{\citenamefont {Eichmann}\ \emph
  {et~al.}(2008{\natexlab{a}})\citenamefont {Eichmann}, \citenamefont
  {Krassnigg}, \citenamefont {Schwinzerl},\ and\ \citenamefont
  {Alkofer}}]{Eichmann:2007nn}%
  \BibitemOpen
  \bibfield  {author} {\bibinfo {author} {\bibfnamefont {G.}~\bibnamefont
  {Eichmann}}, \bibinfo {author} {\bibfnamefont {A.}~\bibnamefont {Krassnigg}},
  \bibinfo {author} {\bibfnamefont {M.}~\bibnamefont {Schwinzerl}}, \ and\
  \bibinfo {author} {\bibfnamefont {R.}~\bibnamefont {Alkofer}},\ }\href
  {\doibase 10.1016/j.aop.2008.02.007} {\bibfield  {journal} {\bibinfo
  {journal} {Annals Phys.}\ }\textbf {\bibinfo {volume} {323}},\ \bibinfo
  {pages} {2505} (\bibinfo {year} {2008}{\natexlab{a}})}\BibitemShut {NoStop}%
\bibitem [{\citenamefont {Eichmann}\ \emph
  {et~al.}(2008{\natexlab{b}})\citenamefont {Eichmann}, \citenamefont
  {Alkofer}, \citenamefont {Cloet}, \citenamefont {Krassnigg},\ and\
  \citenamefont {Roberts}}]{Eichmann:2008ae}%
  \BibitemOpen
  \bibfield  {author} {\bibinfo {author} {\bibfnamefont {G.}~\bibnamefont
  {Eichmann}}, \bibinfo {author} {\bibfnamefont {R.}~\bibnamefont {Alkofer}},
  \bibinfo {author} {\bibfnamefont {I.~C.}\ \bibnamefont {Cloet}}, \bibinfo
  {author} {\bibfnamefont {A.}~\bibnamefont {Krassnigg}}, \ and\ \bibinfo
  {author} {\bibfnamefont {C.~D.}\ \bibnamefont {Roberts}},\ }\href {\doibase
  10.1103/PhysRevC.77.042202} {\bibfield  {journal} {\bibinfo  {journal} {Phys.
  Rev.}\ }\textbf {\bibinfo {volume} {C77}},\ \bibinfo {pages} {042202}
  (\bibinfo {year} {2008}{\natexlab{b}})}\BibitemShut {NoStop}%
\bibitem [{\citenamefont {Eichmann}\ \emph {et~al.}(2009)\citenamefont
  {Eichmann}, \citenamefont {Cloet}, \citenamefont {Alkofer}, \citenamefont
  {Krassnigg},\ and\ \citenamefont {Roberts}}]{Eichmann:2008ef}%
  \BibitemOpen
  \bibfield  {author} {\bibinfo {author} {\bibfnamefont {G.}~\bibnamefont
  {Eichmann}}, \bibinfo {author} {\bibfnamefont {I.~C.}\ \bibnamefont {Cloet}},
  \bibinfo {author} {\bibfnamefont {R.}~\bibnamefont {Alkofer}}, \bibinfo
  {author} {\bibfnamefont {A.}~\bibnamefont {Krassnigg}}, \ and\ \bibinfo
  {author} {\bibfnamefont {C.~D.}\ \bibnamefont {Roberts}},\ }\href {\doibase
  10.1103/PhysRevC.79.012202} {\bibfield  {journal} {\bibinfo  {journal} {Phys.
  Rev.}\ }\textbf {\bibinfo {volume} {C79}},\ \bibinfo {pages} {012202}
  (\bibinfo {year} {2009})}\BibitemShut {NoStop}%
\bibitem [{\citenamefont {Nicmorus}\ \emph {et~al.}(2009)\citenamefont
  {Nicmorus}, \citenamefont {Eichmann}, \citenamefont {Krassnigg},\ and\
  \citenamefont {Alkofer}}]{Nicmorus:2008vb}%
  \BibitemOpen
  \bibfield  {author} {\bibinfo {author} {\bibfnamefont {D.}~\bibnamefont
  {Nicmorus}}, \bibinfo {author} {\bibfnamefont {G.}~\bibnamefont {Eichmann}},
  \bibinfo {author} {\bibfnamefont {A.}~\bibnamefont {Krassnigg}}, \ and\
  \bibinfo {author} {\bibfnamefont {R.}~\bibnamefont {Alkofer}},\ }\href
  {\doibase 10.1103/PhysRevD.80.054028} {\bibfield  {journal} {\bibinfo
  {journal} {Phys. Rev.}\ }\textbf {\bibinfo {volume} {D80}},\ \bibinfo {pages}
  {054028} (\bibinfo {year} {2009})}\BibitemShut {NoStop}%
\bibitem [{\citenamefont {Nicmorus}\ \emph {et~al.}(2010)\citenamefont
  {Nicmorus}, \citenamefont {Eichmann},\ and\ \citenamefont
  {Alkofer}}]{Nicmorus:2010sd}%
  \BibitemOpen
  \bibfield  {author} {\bibinfo {author} {\bibfnamefont {D.}~\bibnamefont
  {Nicmorus}}, \bibinfo {author} {\bibfnamefont {G.}~\bibnamefont {Eichmann}},
  \ and\ \bibinfo {author} {\bibfnamefont {R.}~\bibnamefont {Alkofer}},\ }\href
  {\doibase 10.1103/PhysRevD.82.114017} {\bibfield  {journal} {\bibinfo
  {journal} {Phys. Rev.}\ }\textbf {\bibinfo {volume} {D82}},\ \bibinfo {pages}
  {114017} (\bibinfo {year} {2010})}\BibitemShut {NoStop}%
\bibitem [{\citenamefont {Eichmann}\ \emph
  {et~al.}(2010{\natexlab{b}})\citenamefont {Eichmann}, \citenamefont
  {Alkofer}, \citenamefont {Fischer}, \citenamefont {Krassnigg},\ and\
  \citenamefont {Nicmorus}}]{Eichmann:2010je}%
  \BibitemOpen
  \bibfield  {author} {\bibinfo {author} {\bibfnamefont {G.}~\bibnamefont
  {Eichmann}}, \bibinfo {author} {\bibfnamefont {R.}~\bibnamefont {Alkofer}},
  \bibinfo {author} {\bibfnamefont {C.~S.}\ \bibnamefont {Fischer}}, \bibinfo
  {author} {\bibfnamefont {A.}~\bibnamefont {Krassnigg}}, \ and\ \bibinfo
  {author} {\bibfnamefont {D.}~\bibnamefont {Nicmorus}},\ }\href@noop {} {\
  }\Eprint {http://arxiv.org/abs/1010.0206} {1010.0206 [hep-ph]} \BibitemShut
  {NoStop}%
\bibitem [{\citenamefont {Nicmorus}\ \emph {et~al.}(2011)\citenamefont
  {Nicmorus}, \citenamefont {Eichmann}, \citenamefont {Krassnigg},\ and\
  \citenamefont {Alkofer}}]{Nicmorus:2010mc}%
  \BibitemOpen
  \bibfield  {author} {\bibinfo {author} {\bibfnamefont {D.}~\bibnamefont
  {Nicmorus}}, \bibinfo {author} {\bibfnamefont {G.}~\bibnamefont {Eichmann}},
  \bibinfo {author} {\bibfnamefont {A.}~\bibnamefont {Krassnigg}}, \ and\
  \bibinfo {author} {\bibfnamefont {R.}~\bibnamefont {Alkofer}},\ }\href
  {\doibase 10.1007/s00601-010-0194-5} {\bibfield  {journal} {\bibinfo
  {journal} {Few Body Syst.}\ }\textbf {\bibinfo {volume} {49}},\ \bibinfo
  {pages} {255} (\bibinfo {year} {2011})}\BibitemShut {NoStop}%
\bibitem [{\citenamefont {Eichmann}(2009)}]{Eichmann:2009zx}%
  \BibitemOpen
  \bibfield  {author} {\bibinfo {author} {\bibfnamefont {G.}~\bibnamefont
  {Eichmann}},\ }\href@noop {} {\ }\bibinfo {note} {PhD thesis, University of
  Graz},\ \Eprint {http://arxiv.org/abs/0909.0703} {0909.0703 [hep-ph]}
  \BibitemShut {NoStop}%
\bibitem [{\citenamefont {Eichmann}\ \emph
  {et~al.}(2010{\natexlab{c}})\citenamefont {Eichmann}, \citenamefont
  {Alkofer}, \citenamefont {Krassnigg},\ and\ \citenamefont
  {Nicmorus}}]{Eichmann:2009qa}%
  \BibitemOpen
  \bibfield  {author} {\bibinfo {author} {\bibfnamefont {G.}~\bibnamefont
  {Eichmann}}, \bibinfo {author} {\bibfnamefont {R.}~\bibnamefont {Alkofer}},
  \bibinfo {author} {\bibfnamefont {A.}~\bibnamefont {Krassnigg}}, \ and\
  \bibinfo {author} {\bibfnamefont {D.}~\bibnamefont {Nicmorus}},\ }\href
  {\doibase 10.1103/PhysRevLett.104.201601} {\bibfield  {journal} {\bibinfo
  {journal} {Phys. Rev. Lett.}\ }\textbf {\bibinfo {volume} {104}},\ \bibinfo
  {pages} {201601} (\bibinfo {year} {2010}{\natexlab{c}})}\BibitemShut
  {NoStop}%
\bibitem [{\citenamefont {Alkofer}\ \emph {et~al.}(2010)\citenamefont
  {Alkofer}, \citenamefont {Eichmann}, \citenamefont {Krassnigg},\ and\
  \citenamefont {Nicmorus}}]{Alkofer:2009jk}%
  \BibitemOpen
  \bibfield  {author} {\bibinfo {author} {\bibfnamefont {R.}~\bibnamefont
  {Alkofer}}, \bibinfo {author} {\bibfnamefont {G.}~\bibnamefont {Eichmann}},
  \bibinfo {author} {\bibfnamefont {A.}~\bibnamefont {Krassnigg}}, \ and\
  \bibinfo {author} {\bibfnamefont {D.}~\bibnamefont {Nicmorus}},\ }\href
  {\doibase 10.1088/1674-1137/34/9/005} {\bibfield  {journal} {\bibinfo
  {journal} {Chin. Phys.}\ }\textbf {\bibinfo {volume} {C34}},\ \bibinfo
  {pages} {1175} (\bibinfo {year} {2010})}\BibitemShut {NoStop}%
\bibitem [{\citenamefont {Eichmann}(2011)}]{Eichmann:2011vu}%
  \BibitemOpen
  \bibfield  {author} {\bibinfo {author} {\bibfnamefont {G.}~\bibnamefont
  {Eichmann}},\ }\href {\doibase 10.1103/PhysRevD.84.014014} {\bibfield
  {journal} {\bibinfo  {journal} {Phys. Rev.}\ }\textbf {\bibinfo {volume}
  {D84}},\ \bibinfo {pages} {014014} (\bibinfo {year} {2011})}\BibitemShut
  {NoStop}%
\bibitem [{\citenamefont {Sanchis-Alepuz}\ \emph {et~al.}(2010)\citenamefont
  {Sanchis-Alepuz}, \citenamefont {Alkofer}, \citenamefont {Eichmann},\ and\
  \citenamefont {Villalba-Chavez}}]{SanchisAlepuz:2010in}%
  \BibitemOpen
  \bibfield  {author} {\bibinfo {author} {\bibfnamefont {H.}~\bibnamefont
  {Sanchis-Alepuz}}, \bibinfo {author} {\bibfnamefont {R.}~\bibnamefont
  {Alkofer}}, \bibinfo {author} {\bibfnamefont {G.}~\bibnamefont {Eichmann}}, \
  and\ \bibinfo {author} {\bibfnamefont {S.}~\bibnamefont {Villalba-Chavez}},\
  }\href {http://pos.sissa.it//archive/conferences/119/018/LC2010_018.pdf}
  {\bibfield  {journal} {\bibinfo  {journal} {PoS}\ }\textbf {\bibinfo {volume}
  {LC2010}},\ \bibinfo {pages} {018} (\bibinfo {year} {2010})}\BibitemShut
  {NoStop}%
\bibitem [{\citenamefont {Delbourgo}\ and\ \citenamefont
  {Scadron}(1980)}]{Delbourgo:1979pt}%
  \BibitemOpen
  \bibfield  {author} {\bibinfo {author} {\bibfnamefont {R.}~\bibnamefont
  {Delbourgo}}\ and\ \bibinfo {author} {\bibfnamefont {M.}~\bibnamefont
  {Scadron}},\ }\href {\doibase 10.1088/0305-4616/6/6/003} {\bibfield
  {journal} {\bibinfo  {journal} {J.Phys.G}\ }\textbf {\bibinfo {volume}
  {G6}},\ \bibinfo {pages} {649} (\bibinfo {year} {1980})}\BibitemShut
  {NoStop}%
\bibitem [{\citenamefont {Finger}\ \emph {et~al.}(1980)\citenamefont {Finger},
  \citenamefont {Mandula},\ and\ \citenamefont {Weyers}}]{Finger:1980dw}%
  \BibitemOpen
  \bibfield  {author} {\bibinfo {author} {\bibfnamefont {J.}~\bibnamefont
  {Finger}}, \bibinfo {author} {\bibfnamefont {J.}~\bibnamefont {Mandula}}, \
  and\ \bibinfo {author} {\bibfnamefont {J.}~\bibnamefont {Weyers}},\ }\href
  {\doibase 10.1016/0370-2693(80)90789-3} {\bibfield  {journal} {\bibinfo
  {journal} {Phys.Lett.}\ }\textbf {\bibinfo {volume} {B96}},\ \bibinfo {pages}
  {367} (\bibinfo {year} {1980})}\BibitemShut {NoStop}%
\bibitem [{\citenamefont {Fomin}\ \emph {et~al.}(1983)\citenamefont {Fomin},
  \citenamefont {Gusynin}, \citenamefont {Miransky},\ and\ \citenamefont
  {Sitenko}}]{Fomin:1984tv}%
  \BibitemOpen
  \bibfield  {author} {\bibinfo {author} {\bibfnamefont {P.}~\bibnamefont
  {Fomin}}, \bibinfo {author} {\bibfnamefont {V.}~\bibnamefont {Gusynin}},
  \bibinfo {author} {\bibfnamefont {V.}~\bibnamefont {Miransky}}, \ and\
  \bibinfo {author} {\bibfnamefont {Y.}~\bibnamefont {Sitenko}},\ }\href@noop
  {} {\bibfield  {journal} {\bibinfo  {journal} {Riv.Nuovo Cim.}\ }\textbf
  {\bibinfo {volume} {6N5}},\ \bibinfo {pages} {1} (\bibinfo {year}
  {1983})}\BibitemShut {NoStop}%
\bibitem [{\citenamefont {Munczek}\ and\ \citenamefont
  {Jain}(1992)}]{Munczek:1991jb}%
  \BibitemOpen
  \bibfield  {author} {\bibinfo {author} {\bibfnamefont {H.~J.}\ \bibnamefont
  {Munczek}}\ and\ \bibinfo {author} {\bibfnamefont {P.}~\bibnamefont {Jain}},\
  }\href {\doibase 10.1103/PhysRevD.46.438} {\bibfield  {journal} {\bibinfo
  {journal} {Phys.Rev.}\ }\textbf {\bibinfo {volume} {D46}},\ \bibinfo {pages}
  {438} (\bibinfo {year} {1992})}\BibitemShut {NoStop}%
\bibitem [{\citenamefont {Munczek}(1995)}]{Munczek:1994zz}%
  \BibitemOpen
  \bibfield  {author} {\bibinfo {author} {\bibfnamefont {H.}~\bibnamefont
  {Munczek}},\ }\href {\doibase 10.1103/PhysRevD.52.4736} {\bibfield  {journal}
  {\bibinfo  {journal} {Phys.Rev.}\ }\textbf {\bibinfo {volume} {D52}},\
  \bibinfo {pages} {4736} (\bibinfo {year} {1995})}\BibitemShut {NoStop}%
\bibitem [{\citenamefont {Maris}\ \emph {et~al.}(1998)\citenamefont {Maris},
  \citenamefont {Roberts},\ and\ \citenamefont {Tandy}}]{Maris:1997hd}%
  \BibitemOpen
  \bibfield  {author} {\bibinfo {author} {\bibfnamefont {P.}~\bibnamefont
  {Maris}}, \bibinfo {author} {\bibfnamefont {C.~D.}\ \bibnamefont {Roberts}},
  \ and\ \bibinfo {author} {\bibfnamefont {P.~C.}\ \bibnamefont {Tandy}},\
  }\href {\doibase 10.1016/S0370-2693(97)01535-9} {\bibfield  {journal}
  {\bibinfo  {journal} {Phys. Lett.}\ }\textbf {\bibinfo {volume} {B420}},\
  \bibinfo {pages} {267} (\bibinfo {year} {1998})}\BibitemShut {NoStop}%
\bibitem [{\citenamefont {Holl}\ \emph {et~al.}(2004)\citenamefont {Holl},
  \citenamefont {Krassnigg},\ and\ \citenamefont {Roberts}}]{Holl:2004fr}%
  \BibitemOpen
  \bibfield  {author} {\bibinfo {author} {\bibfnamefont {A.}~\bibnamefont
  {Holl}}, \bibinfo {author} {\bibfnamefont {A.}~\bibnamefont {Krassnigg}}, \
  and\ \bibinfo {author} {\bibfnamefont {C.~D.}\ \bibnamefont {Roberts}},\
  }\href {\doibase 10.1103/PhysRevC.70.042203} {\bibfield  {journal} {\bibinfo
  {journal} {Phys. Rev.}\ }\textbf {\bibinfo {volume} {C70}},\ \bibinfo {pages}
  {042203} (\bibinfo {year} {2004})}\BibitemShut {NoStop}%
\bibitem [{\citenamefont {Maris}\ and\ \citenamefont
  {Tandy}(1999)}]{Maris:1999nt}%
  \BibitemOpen
  \bibfield  {author} {\bibinfo {author} {\bibfnamefont {P.}~\bibnamefont
  {Maris}}\ and\ \bibinfo {author} {\bibfnamefont {P.~C.}\ \bibnamefont
  {Tandy}},\ }\href {\doibase 10.1103/PhysRevC.60.055214} {\bibfield  {journal}
  {\bibinfo  {journal} {Phys. Rev.}\ }\textbf {\bibinfo {volume} {C60}},\
  \bibinfo {pages} {055214} (\bibinfo {year} {1999})}\BibitemShut {NoStop}%
\bibitem [{\citenamefont {Alexandrou}\ \emph {et~al.}(2009)\citenamefont
  {Alexandrou} \emph {et~al.}}]{Alexandrou:2009hs}%
  \BibitemOpen
  \bibfield  {author} {\bibinfo {author} {\bibfnamefont {C.}~\bibnamefont
  {Alexandrou}} \emph {et~al.},\ }\href {\doibase
  10.1016/j.nuclphysa.2009.04.005} {\bibfield  {journal} {\bibinfo  {journal}
  {Nucl. Phys.}\ }\textbf {\bibinfo {volume} {A825}},\ \bibinfo {pages} {115}
  (\bibinfo {year} {2009})}\BibitemShut {NoStop}%
\bibitem [{\citenamefont {Engel}\ \emph {et~al.}(2010)\citenamefont {Engel},
  \citenamefont {Lang}, \citenamefont {Limmer}, \citenamefont {Mohler},\ and\
  \citenamefont {Schafer}}]{Engel:2010my}%
  \BibitemOpen
  \bibfield  {author} {\bibinfo {author} {\bibfnamefont {G.~P.}\ \bibnamefont
  {Engel}}, \bibinfo {author} {\bibfnamefont {C.~B.}\ \bibnamefont {Lang}},
  \bibinfo {author} {\bibfnamefont {M.}~\bibnamefont {Limmer}}, \bibinfo
  {author} {\bibfnamefont {D.}~\bibnamefont {Mohler}}, \ and\ \bibinfo {author}
  {\bibfnamefont {A.}~\bibnamefont {Schafer}} (\bibinfo {collaboration} {BGR
  [Bern-Graz-Regensburg]}),\ }\href {\doibase 10.1103/PhysRevD.82.034505}
  {\bibfield  {journal} {\bibinfo  {journal} {Phys. Rev.}\ }\textbf {\bibinfo
  {volume} {D82}},\ \bibinfo {pages} {034505} (\bibinfo {year}
  {2010})}\BibitemShut {NoStop}%
\bibitem [{\citenamefont {Zanotti}\ \emph {et~al.}(2003)\citenamefont {Zanotti}
  \emph {et~al.}}]{Zanotti:2003fx}%
  \BibitemOpen
  \bibfield  {author} {\bibinfo {author} {\bibfnamefont {J.~M.}\ \bibnamefont
  {Zanotti}} \emph {et~al.} (\bibinfo {collaboration} {CSSM Lattice}),\ }\href
  {\doibase 10.1103/PhysRevD.68.054506} {\bibfield  {journal} {\bibinfo
  {journal} {Phys. Rev.}\ }\textbf {\bibinfo {volume} {D68}},\ \bibinfo {pages}
  {054506} (\bibinfo {year} {2003})}\BibitemShut {NoStop}%
\bibitem [{\citenamefont {Gattringer}\ \emph {et~al.}(2009)\citenamefont
  {Gattringer} \emph {et~al.}}]{Gattringer:2008vj}%
  \BibitemOpen
  \bibfield  {author} {\bibinfo {author} {\bibfnamefont {C.}~\bibnamefont
  {Gattringer}} \emph {et~al.},\ }\href {\doibase 10.1103/PhysRevD.79.054501}
  {\bibfield  {journal} {\bibinfo  {journal} {Phys. Rev.}\ }\textbf {\bibinfo
  {volume} {D79}},\ \bibinfo {pages} {054501} (\bibinfo {year}
  {2009})}\BibitemShut {NoStop}%
\bibitem [{\citenamefont {Fischer}\ and\ \citenamefont
  {Williams}(2008)}]{Fischer:2008wy}%
  \BibitemOpen
  \bibfield  {author} {\bibinfo {author} {\bibfnamefont {C.~S.}\ \bibnamefont
  {Fischer}}\ and\ \bibinfo {author} {\bibfnamefont {R.}~\bibnamefont
  {Williams}},\ }\href {\doibase 10.1103/PhysRevD.78.074006} {\bibfield
  {journal} {\bibinfo  {journal} {Phys. Rev.}\ }\textbf {\bibinfo {volume}
  {D78}},\ \bibinfo {pages} {074006} (\bibinfo {year} {2008})}\BibitemShut
  {NoStop}%
\bibitem [{\citenamefont {Fischer}\ and\ \citenamefont
  {Williams}(2009)}]{Fischer:2009jm}%
  \BibitemOpen
  \bibfield  {author} {\bibinfo {author} {\bibfnamefont {C.~S.}\ \bibnamefont
  {Fischer}}\ and\ \bibinfo {author} {\bibfnamefont {R.}~\bibnamefont
  {Williams}},\ }\href {\doibase 10.1103/PhysRevLett.103.122001} {\bibfield
  {journal} {\bibinfo  {journal} {Phys. Rev. Lett.}\ }\textbf {\bibinfo
  {volume} {103}},\ \bibinfo {pages} {122001} (\bibinfo {year}
  {2009})}\BibitemShut {NoStop}%
\bibitem [{\citenamefont {Chang}\ and\ \citenamefont
  {Roberts}(2009)}]{Chang:2009zb}%
  \BibitemOpen
  \bibfield  {author} {\bibinfo {author} {\bibfnamefont {L.}~\bibnamefont
  {Chang}}\ and\ \bibinfo {author} {\bibfnamefont {C.~D.}\ \bibnamefont
  {Roberts}},\ }\href {\doibase 10.1103/PhysRevLett.103.081601} {\bibfield
  {journal} {\bibinfo  {journal} {Phys. Rev. Lett.}\ }\textbf {\bibinfo
  {volume} {103}},\ \bibinfo {pages} {081601} (\bibinfo {year}
  {2009})}\BibitemShut {NoStop}%
\bibitem [{\citenamefont {Chang}\ \emph
  {et~al.}(2011{\natexlab{b}})\citenamefont {Chang}, \citenamefont {Liu},\ and\
  \citenamefont {Roberts}}]{Chang:2010hb}%
  \BibitemOpen
  \bibfield  {author} {\bibinfo {author} {\bibfnamefont {L.}~\bibnamefont
  {Chang}}, \bibinfo {author} {\bibfnamefont {Y.-X.}\ \bibnamefont {Liu}}, \
  and\ \bibinfo {author} {\bibfnamefont {C.~D.}\ \bibnamefont {Roberts}},\
  }\href {\doibase 10.1103/PhysRevLett.106.072001} {\bibfield  {journal}
  {\bibinfo  {journal} {Phys.Rev.Lett.}\ }\textbf {\bibinfo {volume} {106}},\
  \bibinfo {pages} {072001} (\bibinfo {year} {2011}{\natexlab{b}})}\BibitemShut
  {NoStop}%
\bibitem [{\citenamefont {Qin}\ \emph {et~al.}(2011)\citenamefont {Qin},
  \citenamefont {Chang}, \citenamefont {Liu}, \citenamefont {Roberts},\ and\
  \citenamefont {Wilson}}]{Qin:2011dd}%
  \BibitemOpen
  \bibfield  {author} {\bibinfo {author} {\bibfnamefont {S.-x.}\ \bibnamefont
  {Qin}}, \bibinfo {author} {\bibfnamefont {L.}~\bibnamefont {Chang}}, \bibinfo
  {author} {\bibfnamefont {Y.-x.}\ \bibnamefont {Liu}}, \bibinfo {author}
  {\bibfnamefont {C.~D.}\ \bibnamefont {Roberts}}, \ and\ \bibinfo {author}
  {\bibfnamefont {D.~J.}\ \bibnamefont {Wilson}},\ }\href@noop {} {\ }\Eprint
  {http://arxiv.org/abs/1108.0603} {1108.0603 [nucl-th]} \BibitemShut {NoStop}%
\bibitem [{\citenamefont {Allton}\ \emph {et~al.}(2006)\citenamefont {Allton},
  \citenamefont {Armour}, \citenamefont {Leinweber}, \citenamefont {Thomas},\
  and\ \citenamefont {Young}}]{Allton:2005vm}%
  \BibitemOpen
  \bibfield  {author} {\bibinfo {author} {\bibfnamefont {C.~R.}\ \bibnamefont
  {Allton}}, \bibinfo {author} {\bibfnamefont {W.}~\bibnamefont {Armour}},
  \bibinfo {author} {\bibfnamefont {D.~B.}\ \bibnamefont {Leinweber}}, \bibinfo
  {author} {\bibfnamefont {A.~W.}\ \bibnamefont {Thomas}}, \ and\ \bibinfo
  {author} {\bibfnamefont {R.~D.}\ \bibnamefont {Young}},\ }\href
  {http://arxiv.org/abs/hep-lat/0511004} {\bibfield  {journal} {\bibinfo
  {journal} {PoS}\ }\textbf {\bibinfo {volume} {LAT2005}},\ \bibinfo {pages}
  {049} (\bibinfo {year} {2006})}\BibitemShut {NoStop}%
\bibitem [{\citenamefont {Mader}\ \emph {et~al.}(2011)\citenamefont {Mader},
  \citenamefont {Eichmann}, \citenamefont {Blank},\ and\ \citenamefont
  {Krassnigg}}]{Mader:2011zf}%
  \BibitemOpen
  \bibfield  {author} {\bibinfo {author} {\bibfnamefont {V.}~\bibnamefont
  {Mader}}, \bibinfo {author} {\bibfnamefont {G.}~\bibnamefont {Eichmann}},
  \bibinfo {author} {\bibfnamefont {M.}~\bibnamefont {Blank}}, \ and\ \bibinfo
  {author} {\bibfnamefont {A.}~\bibnamefont {Krassnigg}},\ }\href {\doibase
  10.1103/PhysRevD.84.034012} {\bibfield  {journal} {\bibinfo  {journal} {Phys.
  Rev.}\ }\textbf {\bibinfo {volume} {D84}},\ \bibinfo {pages} {034012}
  (\bibinfo {year} {2011})}\BibitemShut {NoStop}%
\end{thebibliography}%

\end{document}